\documentclass[twocolumn]{aastex63}

\DeclareRobustCommand{\ion}[2]{%
\relax\ifmmode
\ifx\testbx\f@series
{\mathbf{#1\,\mathsc{#2}}}\else
{\mathrm{#1\,\mathsc{#2}}}\fi
\else\textup{#1\,{\mdseries\textsc{#2}}}%
\fi}

\let\bfseries\mdseries
\received{X N, 2022}
\revised{Y N, 2022}
\accepted{\today}
\submitjournal{ApJ}
\shorttitle{Detection of Two Components in the CGM of the Milky Way}
\shortauthors{Bluem et al.}
\graphicspath{{./}{figures/}}

\begin{document}

\title{Widespread Detection of Two Components in the Hot Circumgalactic Medium of the Milky Way}

\email{jesse-bluem@uiowa.edu}

\author{Jesse Bluem}
\affiliation{University of Iowa Department of Physics and Astronomy, Van Allen Hall, 30 N. Dubuque St., Iowa City, IA 52242, USA}

\author{Philip Kaaret}
\affiliation{University of Iowa Department of Physics and Astronomy, Van Allen Hall, 30 N. Dubuque St., Iowa City, IA 52242, USA}

\author{K. D. Kuntz}
\affiliation{NASA Goddard Space Flight Center, Greenbelt, MD 20771, USA}
\affiliation{The Henry A. Rowland Department of Physics and Astronomy, Johns Hopkins University, 3701 San Martin Dr., Baltimore, MD 21218, USA}

\author{Keith M. Jahoda}
\affiliation{NASA Goddard Space Flight Center, Greenbelt, MD 20771, USA}

\author{Dimitra Koutroumpa}
\affiliation{LATMOS/IPSL, CNRS, UVSQ Universit\'{e} Paris-Sarclay, Sorbonne Universit\'{e}, Guyancourt, France}

\author{Edmund J. Hodges-Kluck}
\affiliation{NASA Goddard Space Flight Center, Greenbelt, MD 20771, USA}

\author{Chase A. Fuller}
\affiliation{University of Iowa Department of Physics and Astronomy, Van Allen Hall, 30 N. Dubuque St., Iowa City, IA 52242, USA}

\author{Daniel M. LaRocca}
\affiliation{University of Iowa Department of Physics and Astronomy, Van Allen Hall, 30 N. Dubuque St., Iowa City, IA 52242, USA}

\author{Anna Zajczyk}
\affiliation{University of Iowa Department of Physics and Astronomy, Van Allen Hall, 30 N. Dubuque St., Iowa City, IA 52242, USA}
\affiliation{NASA Goddard Space Flight Center, Greenbelt, MD 20771, USA}
\affiliation{Center for Space Sciences and Technology, University of Maryland, Baltimore County, 1000 Hilltop Circle, Baltimore, MD 21250, USA} 



\begin{abstract}
Surrounding the Milky Way (MW) is the circumgalactic medium (CGM), an extended reservoir of hot gas that has significant implications for the evolution of the MW. We used the HaloSat all-sky survey to study the CGM's soft X-ray emission in order to better define its distribution and structure. We extend a previous HaloSat study of the southern CGM (Galactic latitude $b < -30\arcdeg$) to include the northern CGM ($b > 30\arcdeg$) and find evidence that at least two hot gas model components at different temperatures are required to produce the observed emission. The cooler component has a typical temperature of kT $\rm \sim 0.18$ keV, while the hotter component has a typical temperature of kT $\rm \sim 0.7$ keV. The emission measure in both the warm and hot components has a wide range ($\rm \sim 0.005-0.03$, $\rm \sim 0.0005-0.004$ $\rm cm^{-6}\,pc$ respectively), indicating that the CGM is clumpy. A patch of relatively consistent CGM was found in the north, allowing for the CGM spectrum to be studied in finer detail using a stacked spectrum. The stacked spectrum is well described with a model including two hot gas components at temperatures of kT = $\rm 0.166 \pm 0.005$ keV and kT = $\rm 0.69^{+0.04}_{-0.05}$ keV. As an alternative to adding a hot component, a neon-enhanced single-temperature model of the CGM was also tested and found to have worse fit statistics and poor residuals.
\end{abstract}



\section{Introduction} \label{sec:intro}

Surrounding the Galactic disk is a low density bubble of plasma and gas often referred to as the halo or circumgalactic medium (CGM, see \citet{Putman2012} and \citet{Tumlinson2017} for reviews). The CGM extends from the Galactic disk out at least as far as the virial radius of the Galaxy ($\rm \sim 250$ kpc) \citep{Klypin2002,Sommer2006}. The CGM is hot ($\rm \sim 0.1$ - $\rm 0.3$ keV) compared to the relatively cool Galactic disk. Even though the density of the CGM is low, the enormous volume it encompasses allows the CGM to contain a significant amount of material. The exact amount of material in the CGM is unknown, but may be as high as $\rm 1.2 \times 10^{11}$ $\rm M_{\odot}$ \citep{Nicastro2016,Faerman2017}. This is a similar to the total mass of stars in the Milky Way of $6.43 \pm 0.63 \times 10^{10}$ $\rm M_{\odot}$ \citep{McMillan2011}. The CGM mass may provide enough baryons to explain the so-called ``missing baryon problem" (see \citet{Sommer2006}, \citet{Faerman2017}). On the other hand, some mass estimates suggest otherwise; \citet{Miller2015} find an estimate of Galactic mass of $\rm 1.2^{+0.9}_{-0.8} \times 10^{10}$ $\rm M_{\odot}$ out to 250 kpc, an amount that only accounts for less than half of the missing baryonic mass. \citet{Salem2015} studied the orbit and structure of the Large Magellanic Cloud and estimated the CGM mass as $\rm 2.7 \pm 1.4 \times 10^{10}$ $\rm M_{\odot}$ out to 300 kpc, accounting for only 15\% of the total Galactic baryons. \citet{Kaaret2020} estimates the mass in their observed disk-like component of the CGM as $\rm 2 \times 10^{7}$ $\rm M_{\odot}$, but notes the need for an extended component (unobserved in that work) to match known CGM absorption values, which increases the mass estimate to $\rm 5.5-8.6 \times 10^{10}$ $\rm M_{\odot}$.
 
We know this reservoir of hot gas exists outside the Galactic disk for many reasons. Evidence of its presence can be seen in both X-ray emission and absorption lines. The CGM emission is widespread across the sky \citep{Henley2013}. X-ray absorption measurements support gas outside the Galactic disk being from the CGM rather than from the Local Group medium \citep{Bregman2007b}. Combinations of X-ray emission and absorption point towards the CGM having a radius extending out at least as far as 100 kpc, with a mass greater than $\rm 10^{10}$ $\rm M_{\odot}$ \citep{Gupta2012}.

Additional evidence comes from high velocity clouds (HVCs) that have been detected in the CGM \citep{Muller1963}. HVCs are cool, dense clouds with line of sight velocities that exceed that expected from the rotation of the Galactic disk. These HVCs exhibit evidence of being embedded in a hotter medium, that is responsible for stabilizing these clouds. Without this stabilization effect, these clouds would not have a lifetime long enough for them to reach such high velocities \citep{Konz2002}. The interaction between these cool HVCs and the hotter medium produces a detectable \ion{O}{vi} absorption line \citep{Sembach2003}. HVCs also exhibit a distinctive tadpole shape consistent with these clouds moving through some sort of medium \citep{Bruns2000}. Another observation of the existence of the CGM is that satellite galaxies are noticeably stripped of gas as they pass close to the Milky Way, due to the presence of the CGM material \citep{Grcevich2009,Blitz2000}. This effect is not just seen for the Milky Way, but is also seen in other galaxies as well. However, detection of the CGM is only part of the battle, the processes at play in the CGM must also be studied.

Feedback processes between the Galactic disk and the CGM appear to be very important to understanding the CGM. The disk is the likely source of any metals found in the CGM, as the material that accretes to the CGM from the spaces between galaxies should only be weakly metal enriched. It has been observed that starburst galaxies have an enriched CGM \citep{Heckman2016}, pointing towards star formation producing significant feedback. Stars in the disk act as a heat source for the CGM, changing its ionization state. As much as 2\% of the ionizing radiation produced by stars in the Galactic disk reaches the CGM \citep{Bland1999}. Supernovae and active galactic nuclei similarly heat the CGM and can provide the mechanical energy needed to eject material into the CGM \citep{Veilleux2005}. 

We also know that material must travel from the CGM to the disk. Long term star formation rates in galaxies are inconsistent with instantaneous censuses of available stellar fuel in galaxies - there does not appear to be enough fuel in a galaxy at any given time to cover its entire star formation history \citep{Bigiel2008}. This means additional material must be entering galactic disks in order for star formation to continue at the observed rates \citep{Oser2010}. This balance between in-going and outgoing material is crucial to understanding the CGM.
 
The recent work of \citet{Kaaret2020} studied the southern part of the CGM of the Milky Way with measurements of soft X-ray emission, predominantly \ion{O}{vii} and \ion{O}{viii} emission lines, and found it to be clumpy, with a scale height of 1-2 kpc. The intensity of X-ray emission was found to be correlated with the underlying surface density of molecular hydrogen, which serves as a tracer of star formation. This correlation strongly implies that stellar feedback is powering the CGM's observed emission. On the other hand, if the CGM structure was dictated primarily by accretion from the dispersed intergalactic medium then it should be more evenly distributed around the Galaxy. 
 
Another aspect of the CGM that can be studied is the temperature distribution. While the CGM has typically been observed with a single thermal component around $\rm \sim 0.2$ keV \citep{Kaaret2020,Nakashima2018,Gupta2014}, a secondary hotter CGM component is suggested by the work of \citet{Das2019a,Das2019c}, with detections in both absorption (0.786-1.22 keV) and emission (0.413-0.717 keV) spectra. However, this line of sight was near the Fermi bubbles at the center of the Galaxy, and the bubbles could not be ruled out as a source. The Fermi bubbles are a pair of gamma ray emitting regions observed to the north and south of the Galactic center, potentially connected to activity from the Galaxy's central super-massive black hole (for an overview, see \citet{Su2010}). A second, hotter, temperature component (kT = 0.65-0.90 keV) is also observed in \citet{Gupta2020} in emission spectra, although they cannot distinguish between the second, hotter component and an isothermal model with enhanced neon abundances. The results from \citet{Gupta2020} are consistent with \citet{Das2019a,Das2019c}, with sightlines that are not near the Fermi bubbles. Overall, the addition of a hot component to the CGM spectra is not yet the consensus, since supporting observations only existing for particular sightlines. Some external galaxy CGMs are detected with two temperature components ($\rm \sim 0.2$ and $\rm \sim 0.65$ keV, \citet{Owen2009}, $\rm \sim 0.2$ and $\rm \sim 0.6$ keV, \citet{Li2008}, $\rm \sim 0.06-0.17$ and $\rm \sim 0.19-0.37$ keV, \citet{Tullman2006}, $\rm 0.11^{+0.03}_{-0.02}$ and $\rm 0.37 \pm 0.03$ keV, \citet{Strickland2004}, $\rm \sim 0.3$ and $\rm \sim 0.7-0.8$ keV, \citet{Immler2003}, $\rm \sim 0.20$ and $\rm \sim 0.75$ keV, \citet{Kuntz2003}, $\rm 0.166^{+0.009}_{-0.008}$ and $\rm 0.62 \pm 0.04$ keV, \citet{Wang2003}).

The goals of this paper are to study the CGM emission throughout the northern and southern CGM in finer detail than ever before, using observations from HaloSat, a CubeSat X-ray observatory \citep{Kaaret2019}. The design and mission parameters of HaloSat provide an opportunity to study the large-scale spectral features of the CGM in better detail than XMM Newton, Suzaku, or ROSAT. HaloSat provides a valuable opportunity to study the distribution of CGM material, as well as check for the existence and distribution of any additional thermal components in the CGM spectrum.
 
Section 2 describes HaloSat and the observations used in this analysis. Section 3 details the spectral model used for the individual CGM fields. Section 4 discusses the spectral fitting of the individual fields. Section 5 considers the results of the spectral fits for the individual fields. Section 6 covers a deeper look at the CGM using stacked spectra. Section 7 discusses the results of this paper.
 
 \begin{figure*}[htb!]
\centering
\includegraphics[width=1\textwidth]{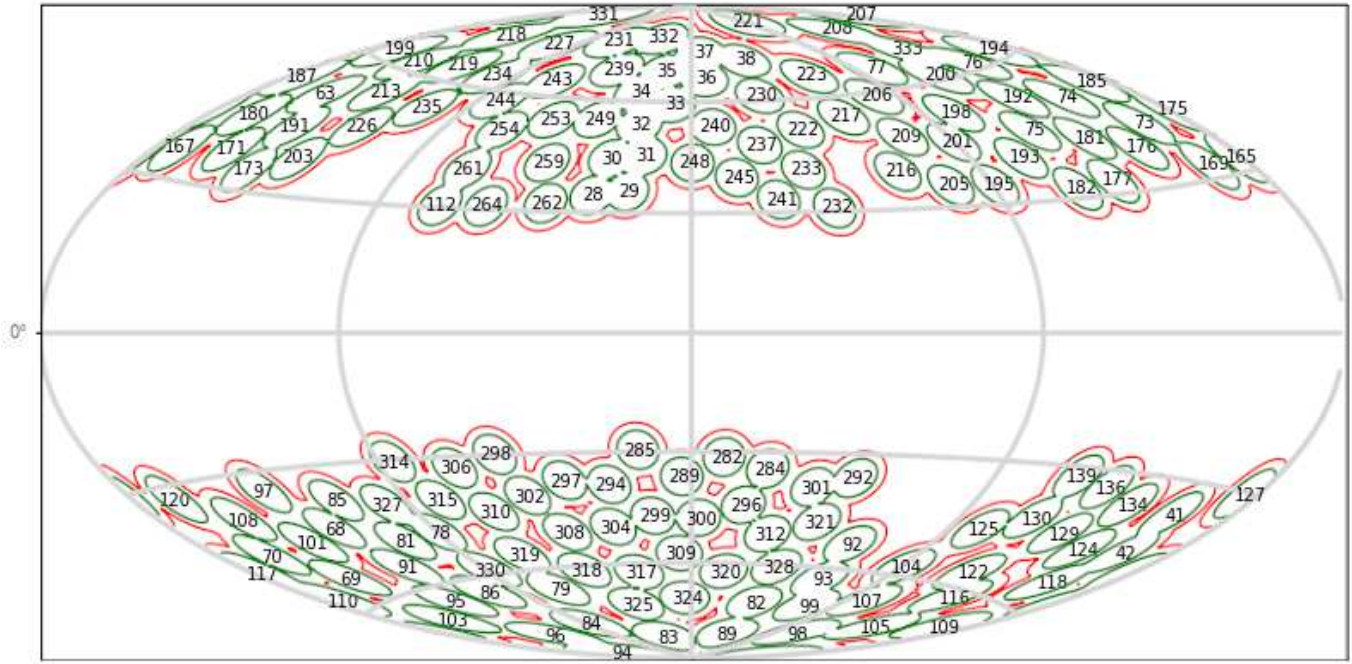}
\caption{HaloSat fields included in this analysis shown in Galactic coordinates. The Galactic center in the middle of the figure. Each field is marked with its HaloSat catalog number. The green contour is the full response field-of-view and the red contour is where the response drops to zero. The gaps in the upper left and lower right are the fields near the ecliptic poles, that were removed by the Sun angle selections.} 
\end{figure*}
 
\section{Observations} \label{sec:obs}

HaloSat was a NASA funded CubeSat with instrumentation developed at the University of Iowa that was designed to detect soft X-rays from extended sources such as the CGM. Understanding the CGM has proven difficult in large part due to the difficulty in observing it. Part of this difficulty stems from the effects of solar wind charge exchange (SWCX). SWCX occurs when solar wind ions and neutral species charge-exchange and the excited end products transition to the ground state, emitting X-ray photons (for a review see \citet{Kuntz2019}). SWCX is problematic because it produces strong, time-variable line emission in some of the same lines used to diagnose the temperature of collisionally ionized plasmas, including the CGM. Disentangling SWCX from these other components has proven difficult. However, the HaloSat observing strategy minimized SWCX contamination, making it ideal for studying the CGM \citep{Kaaret2019}.

Because SWCX is produced by the solar wind, HaloSat minimized the SWCX contamination by taking observations only in the night half of its orbit around the Earth, with the spacecraft observing in directions opposite of the Sun in order to minimize the bright, strongly varying magnetospheric SWCX. The remaining SWCX contribution was modeled and included as components in our spectral fit. This allowed HaloSat to handle SWCX emission much better than previous X-ray instruments.

HaloSat was deployed from the International Space Station on 2018 July 13 at an orbital altitude of $\rm \sim 400$ km \citep{Kaaret2019,LaRocca2020}. HaloSat observed an energy range of 0.4-7.0 keV using three non-imaging silicon drift detectors with an average energy resolution of 85 eV at 676 eV (fluorine K alpha) and 137 eV at 5.9 keV (manganese K alpha), with each detector having an effective area of 5.1 $\rm mm^2$ at 600 eV \citep{Kaaret2019}. HaloSat had full response over a 10 degree diameter field-of-view, which then falls off linearly to zero at 14 degrees diameter. The large field-of-view of HaloSat improves the ratio of signal from diffuse emission to noise from instrumental background, because the instrumental background depends on detector size while the diffuse emission signal increases with field-of-view.

Only the HaloSat fields above Galactic latitudes of $\rm 30^{\circ}$ and below Galactic latitudes of $\rm -30^{\circ}$ were analyzed, in order to avoid the Galactic disk, minimizing the absorption and avoiding contaminating sources and complex regions. Various other selections were performed on the data as well. A Sun angle selection of greater than or equal to 110 degrees was performed to reduce contributions from magnetospheric SWCX and reduce variation from heliospheric SWCX (HSWCX) \citep{Kuntz2019}. Selections were also performed on the standard HaloSat hard band (3.0-7.0 keV) and very large event (VLE) band (7.0+ keV) of 0.12 c/s and 0.75 c/s, respectively. The hard band primarily contains instrumental background counts and the CXB, while the VLE band is out of the nominal data range, so filtering on these background enables us to filter out times when the instrumental background is elevated. For more detail on the instrumental background for HaloSat, see \citet{Kaaret2019}. Two fields with less than 5000 seconds of data remaining after all selections were removed from the data set. An additional complicating factor for the northern CGM is the North Polar Spur (NPS), for an overview, see \citet{LaRocca20202}), which for this analysis was treated as a CGM feature. In the southern hemisphere, the Eridanus Enhancement \citep{Burrows1993} was treated similarly. Two offset fields for the Eridanus Enhancement were also removed from the data set, in favor of keeping the deeper primary observations of those fields. The locations of included fields can be seen in Figure 1.

\begin{figure}[htb!]
\centering
\includegraphics[width=0.48\textwidth]{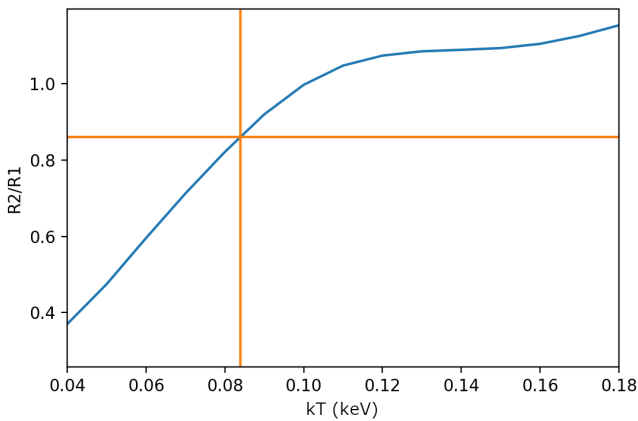}
\caption{The blue line is the ROSAT R2/R1 ratio for a given APEC temperature. The orange lines mark the R2/R1 ratio for the LHB of 0.86, which corresponds to a temperature of kT = 0.084 keV.}
\end{figure}

\section{Spectral Model} \label{sec:ana}
The fitting of individual fields was done with a PyXspec pipeline \citep{Arnaud1996,Gordon2021} and XSPEC version 12.11.1. Spectra were fit over an energy range of 0.4 to 7.0 keV, and each of the three HaloSat detectors was treated as a separate data set.

For each field, the CGM was modeled as a collisionally ionized diffuse gas in equilibrium, using the APEC (Astrophysical Plasma Emission Code) model in XSPEC (Xspec model {\tt apec}) \citep{Smith2001}, with an abundance of 0.3 solar \citep{Bregman2018} and full Galactic absorption particular to each field. The CGM temperature (kT) and emission measure (EM) were free to be fit. All CGM parameters were linked for each HaloSat detector during fitting. Abundances were set to Wilms in Xspec \citep{Wilms2000}. The Wilms abundance table is the basis for Tuebingen-Boulder ISM absorption model (TBabs), the choice of absorption model in analysis of emission from diffuse material. The table is also based on ISM abundances rather than solar, making it more appropriate for this analysis.

All absorptions use the TBabs ISM absorption model (XSPEC model {\tt tbabs}), which calculates the X-ray absorption cross section as a combination of gas, molecular, and grain cross sections in the interstellar medium \citep{Wilms2000}. The full Galactic absorption for each HaloSat field was calculated using Planck dust radiance maps, as radiance serves as a better tracer of the total absorption column density at the relatively low levels of absorption seen towards the Galactic poles when compared to dust opacity \citep{Planck2014}. The radiance maps were first converted to E(B-V) maps following \citet{Planck2014}. The E(B-V) map was then converted to $\rm N_H$ following \citet{Zhu2017}. The best-fit equivalent $\rm N_H$ value was found by fitting combined response-weighted absorption curves over the entire field \citep{LaRocca20202}. This produced a single-valued $\rm N_H$ that more appropriately reflects the range of absorptions over the extended field when compared to a simple average absorption.

In addition to the absorption, there are other important background and foreground astrophysical components in the spectra. These include the cosmic X-ray background (CXB) and local hot bubble (LHB). The CXB was modeled by an absorbed power-law (XSPEC model {\tt powerlaw}), using the photon index and normalization from \citet{Cappelluti2017}, with the normalization adjusted for the HaloSat field-of-view. The absorption value was the same as that used for the CGM. These values were fixed and the same for each detector.

The LHB was included in the model as an unabsorbed plasma in collisional ionization equilibrium. The parameters of the LHB APEC were derived using the method described in \citet{Liu2017}. We have updated that analysis because the existing values from \citet{Liu2017} are no longer appropriate due to updates to the APEC model since that paper's publication date. It is important to note that the AtomDB version used for this analysis was 3.0.9 (July 2020 201 temperature release), and that the choice of absorption table affects the APEC model as well. Since Wilms abundances were used for other aspects of this analysis, it must be used for the LHB calculations as well.

To determine the temperature of the LHB, we can take a series of unabsorbed APEC models for different temperatures, use XSPEC's FakeIt function to turn them into synthetic ROSAT spectra, and determine the ratio of the R2 band to the R1 band in those spectra. We can then determine the real R2/R1 value for the LHB using distance separated ROSAT data based on soft X-ray shadows (see \citet{Snowden1998,Snowden2000}), and find the corresponding temperature. Thus, we find a revised temperature for the LHB of kT = 0.084 keV. This can be seen in Figure 2. The error on this value is likely similar to the error value on the original value from \citet{Liu2017} of $\rm \pm 0.019$ keV. Recreating the derivation of that error for this new LHB temperature is beyond the scope of this paper. It is unlikely to change significantly given that the R2/R1 conversion to EM is fairly linear in the regime of interest in Figure 2. Once we have the LHB's temperature, we can then convert the map of LHB R12 count rates to emission measure (EM), following \citet{Liu2017}. The APEC normalization was calculated using the standard HaloSat vignetting function \citep{LaRocca20202}. This temperature and normalization was then used as the fixed parameters for the LHB APEC and were the same for each detector.

In addition to the cosmic foreground sources, there are HSWCX contributions as well as events in the HaloSat detectors due to the particle background. The HSWCX flux in each HaloSat observation was estimated \citep{Koutroumpa2007} using data from the Solar Wind Ion Composition Spectrometer (SWICS) on the Advanced Composite Explorer (ACE), located at the L1 Lagrange point \citep{Gloeckler1998}. For each HaloSat exposure, we estimated the relevant ACE data that apply to the local ion flux along the line of sight by calculating the propagation time from ACE's position using real-time solar wind speed data. Further detail on this process is described in \citet{Kaaret2020} and \citet{Ringuette2021}. This was included in the model fit as estimates of \ion{O}{vii} and \ion{O}{viii} emission, as a pair of Gaussians (XSPEC model {\tt gauss}), at 0.5634 keV and 0.6531 keV respectively. The HSWCX parameters were the same for each detector.

The local particle backgrounds experienced by the spacecraft were included in the model fit as power-laws (XSPEC model {\tt powerlaw}, without the photon redistribution function) specific to each HaloSat detector. This was because the detectors experienced different energetic charged-particle induced backgrounds due to different locations within the spacecraft. We have adopted a model with two background power-laws, one with a shallower photon index that contributes strongly across all energies, similar to previous HaloSat works, and a steeper secondary component that primarily adds emission at the lowest energies. Using two power-laws to represent the instrumental background provided a significant improvement in fit quality and residuals relative to a single power-law. The shallower power-law was fit across the full 0.4-7.0 keV energy range, simultaneously with the astrophysical model, with photon indexes and normalizations free to fit independently for each detector and with unrestricted ranges. Since the steeper power-law has a weaker contribution to the overall spectrum, its photon indexes were linked between the detectors, while the normalizations were left free to fit independently for each detector and with no restrictions on the range of the parameters. The steeper power-law was also fit simultaneously with the astrophysical model over the full energy range. In addition to fitting individual fields, we also fit stacked spectra coadding several similar CGM fields to obtain a spectra with a much higher signal-to-noise (see Section 6). Due to the high statistics in the stacked spectra, the steeper power-law was fit with a freed photon index. In the individual fields, the photon index was frozen to the stacked spectrum value of 3.4. This was due to the large uncertainty in the photon index in the individual field fits and was the only difference in fitting between the stacked spectra and the individual fields. When the fixed index fits are compared to the freed index fits for the same fields, the difference in fit quality is minor.

For the individual fields, Markov Chain Monte Carlo (MCMC) was used to generate error values for the fits, and as a check on the best fit values being reasonable. The MCMC was run using the Goodman-Weare method \citep{Goodman2010}, with 10 walkers, 15000 steps and a burn length of 1500. Error values (90\% confidence intervals) for the model parameters were generated through the PyXspec error command, using the MCMC chains.

\begin{figure*}[htb!]
\centering
\begin{tabular}{cc}
  \includegraphics[width=0.49\textwidth]{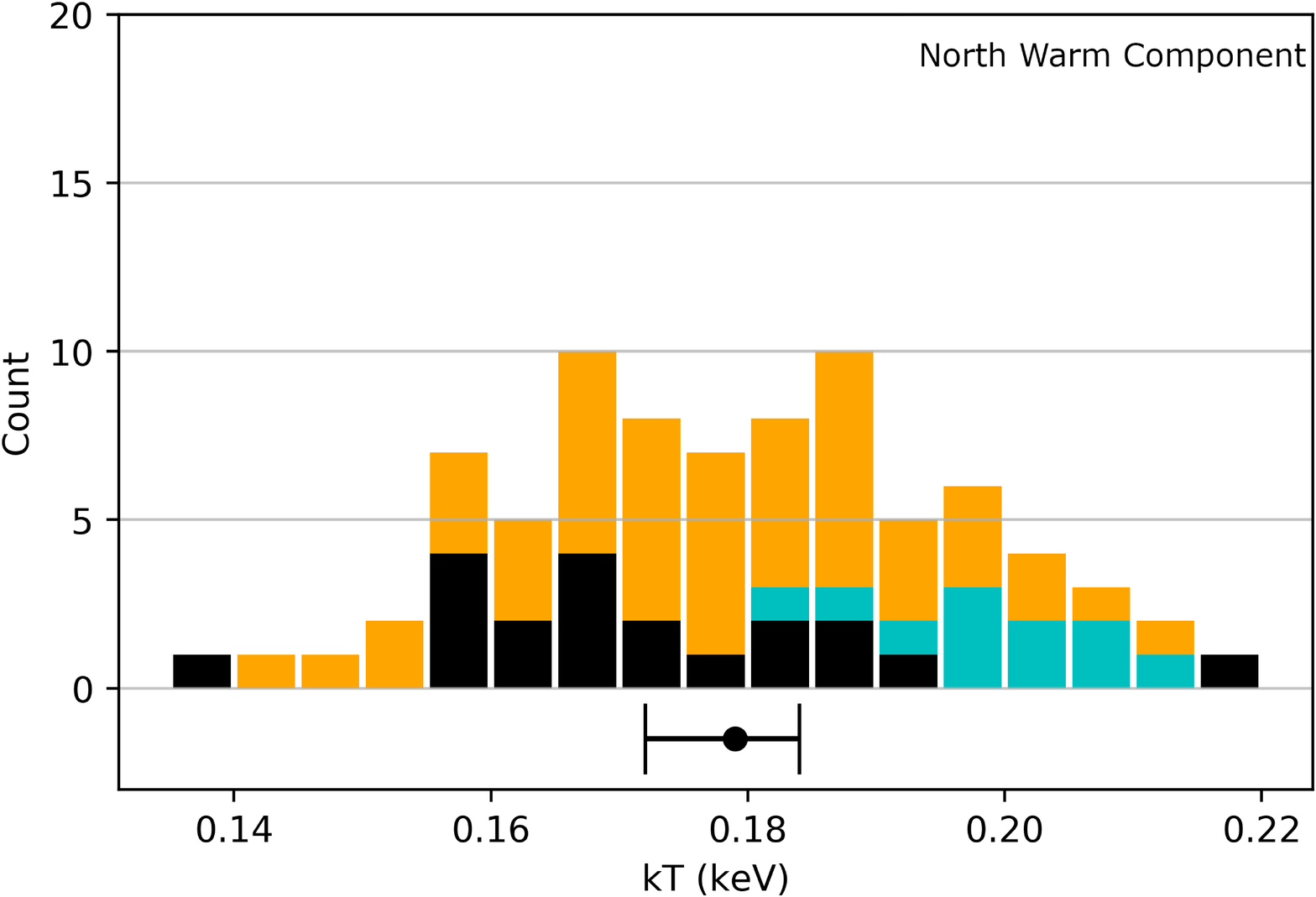} & \includegraphics[width=0.49\textwidth]{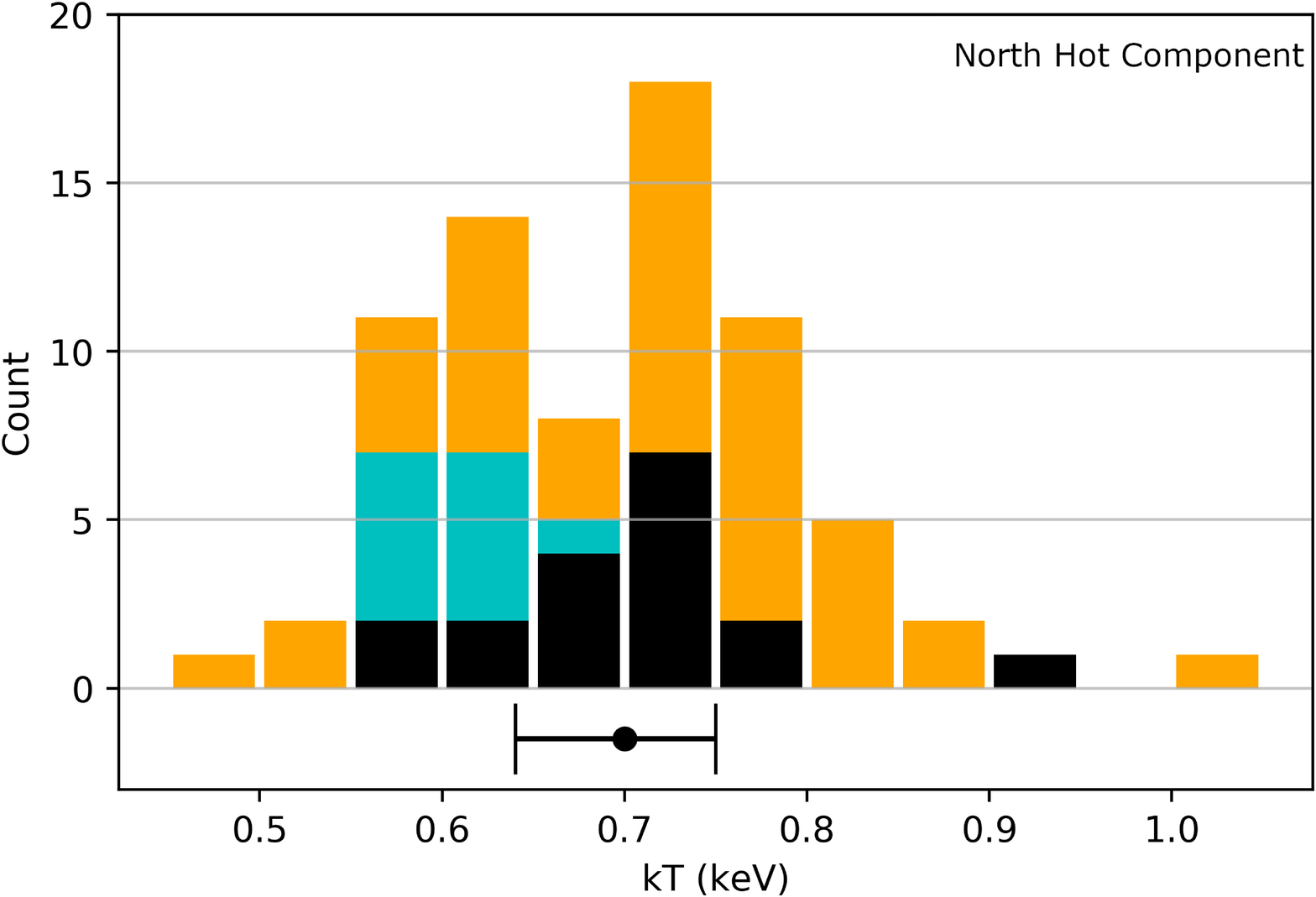} \\
 \includegraphics[width=0.49\textwidth]{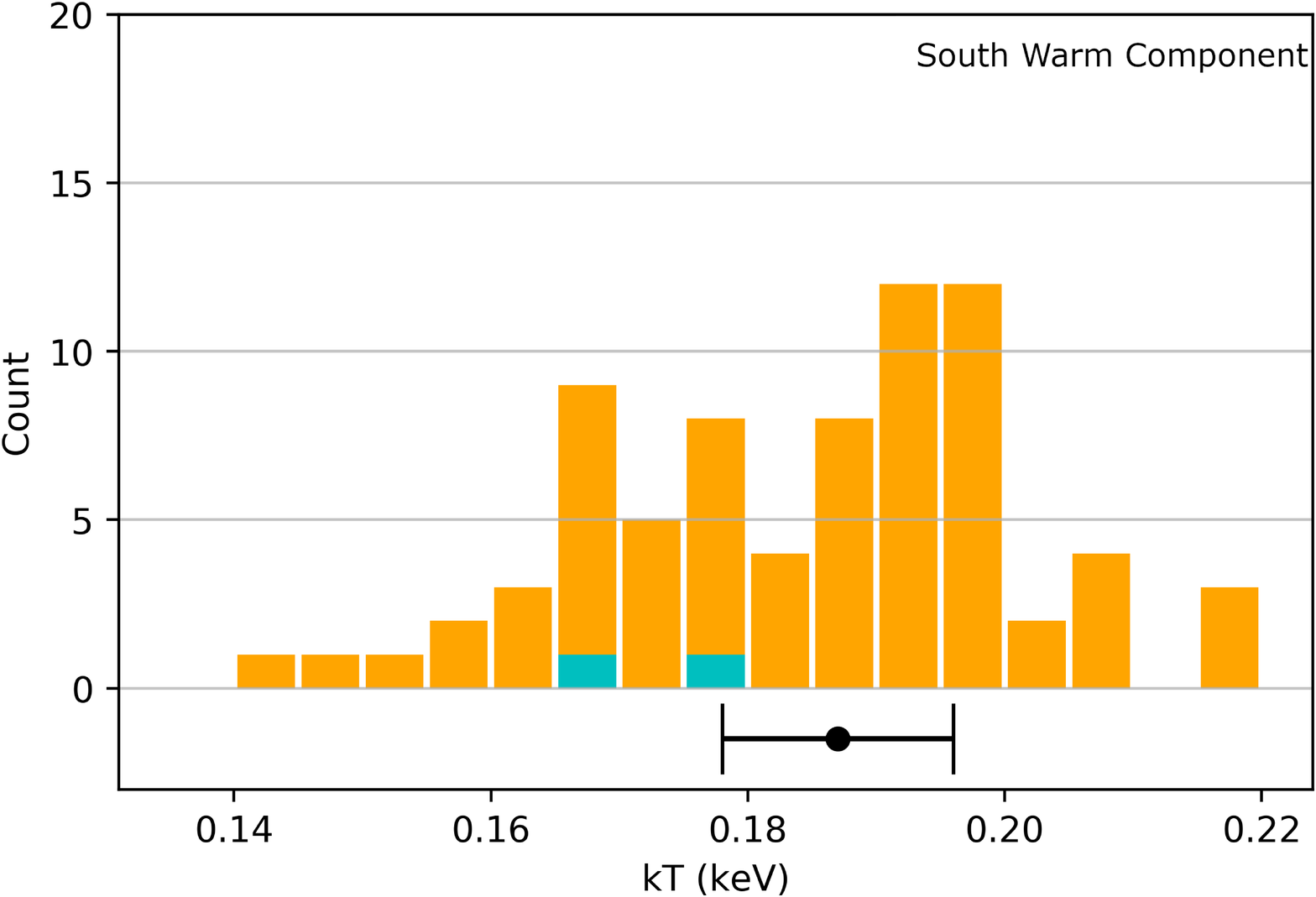} & \includegraphics[width=0.49\textwidth]{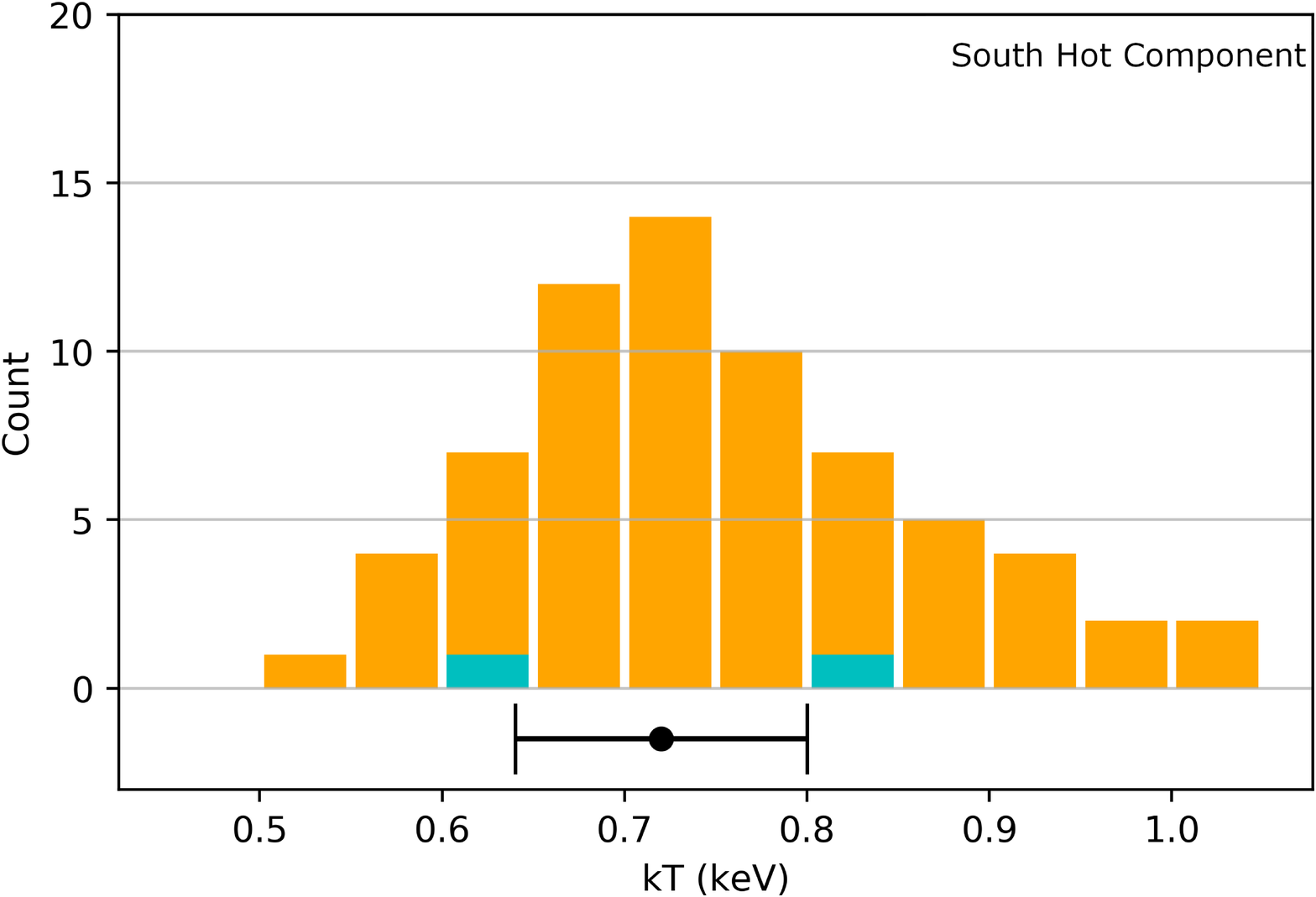} \\
\end{tabular}
\caption{Histograms of temperature for the northern and southern CGM for both the warm and hot components. Marked in cyan are the North Polar Spur for the northern CGM and the Eridanus Enhancement for the southern CGM. Marked in black are the fields used in the stacked spectra. Near the bottom of each histogram is the median value for the data, with an uncertainty that is equivalent to the average uncertainty (90\% confidence interval). This uncertainty serves to compare the typical uncertainty in the individual measurements to the distribution of measured values.} 
\end{figure*}

\begin{figure*}[htb!]
\centering
\begin{tabular}{cc}
  \includegraphics[width=0.49\textwidth]{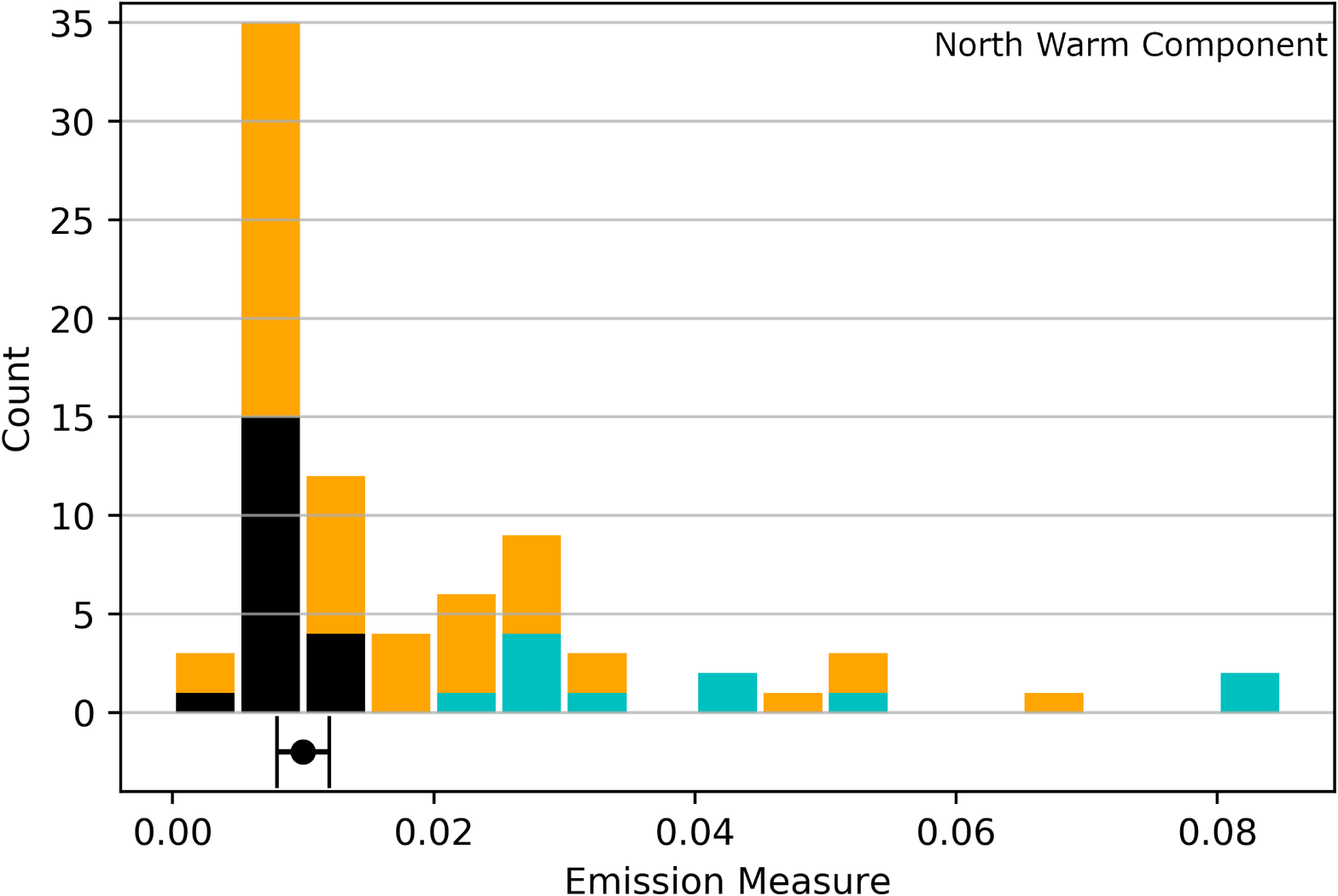} & \includegraphics[width=0.49\textwidth]{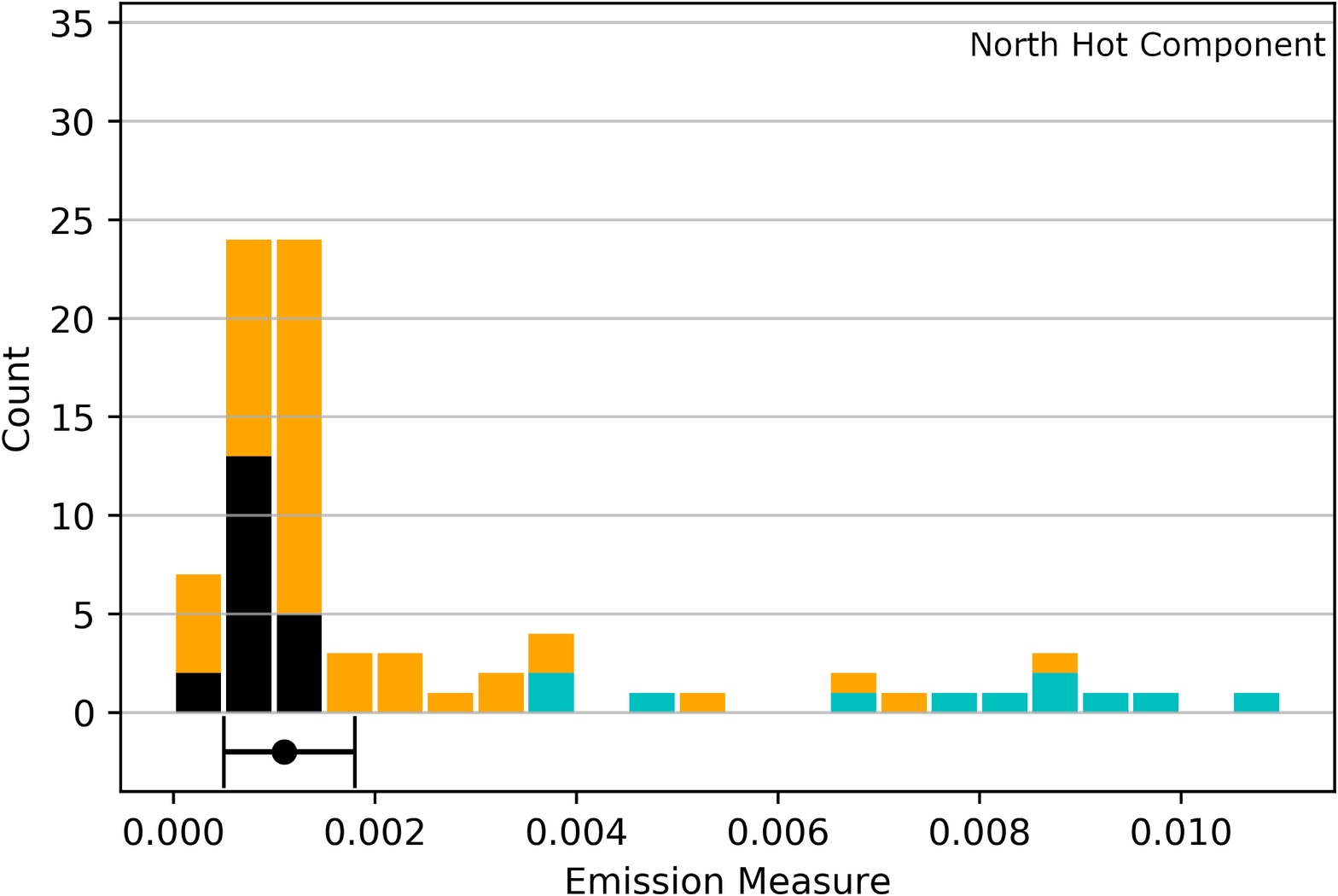} \\
 \includegraphics[width=0.49\textwidth]{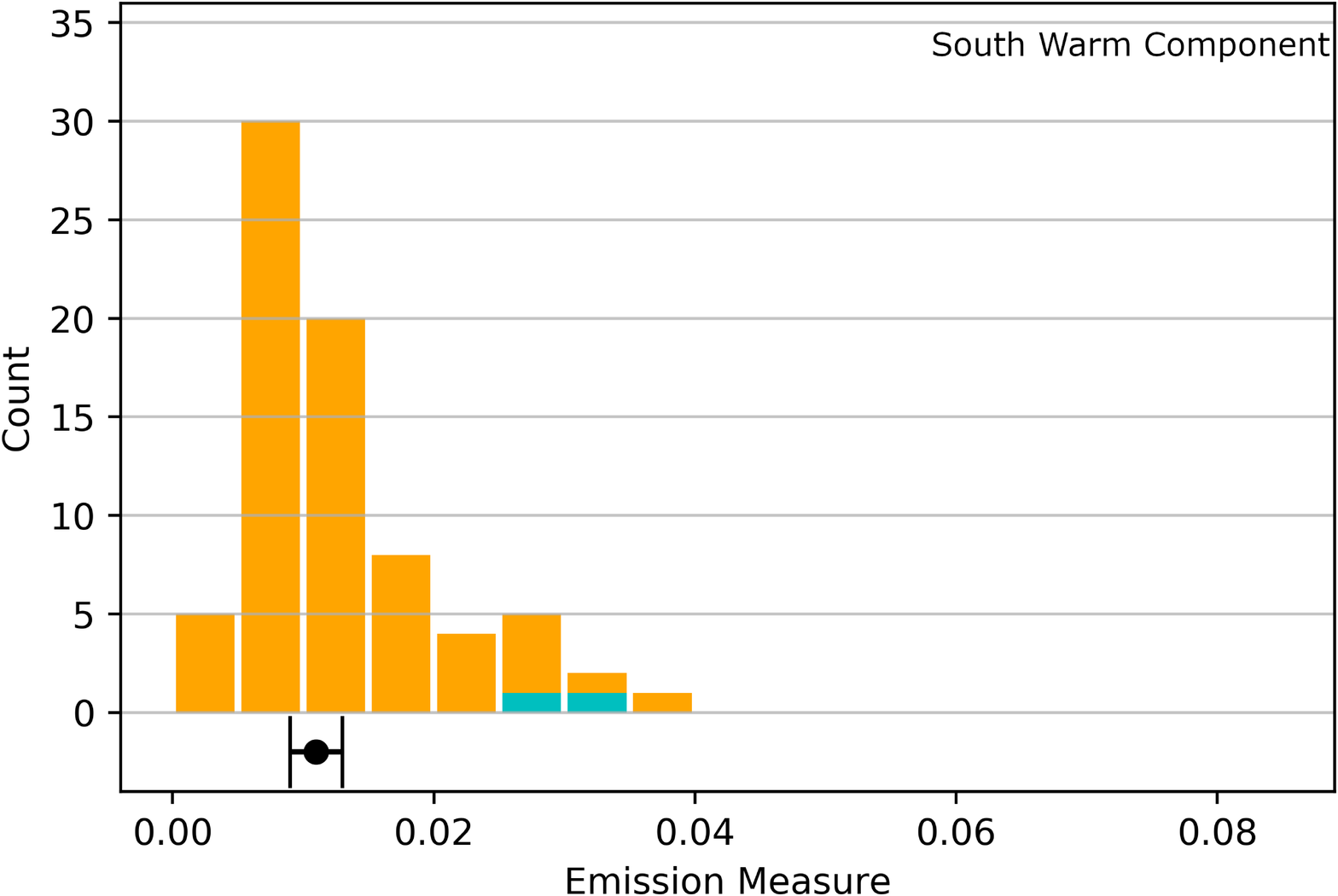} & \includegraphics[width=0.49\textwidth]{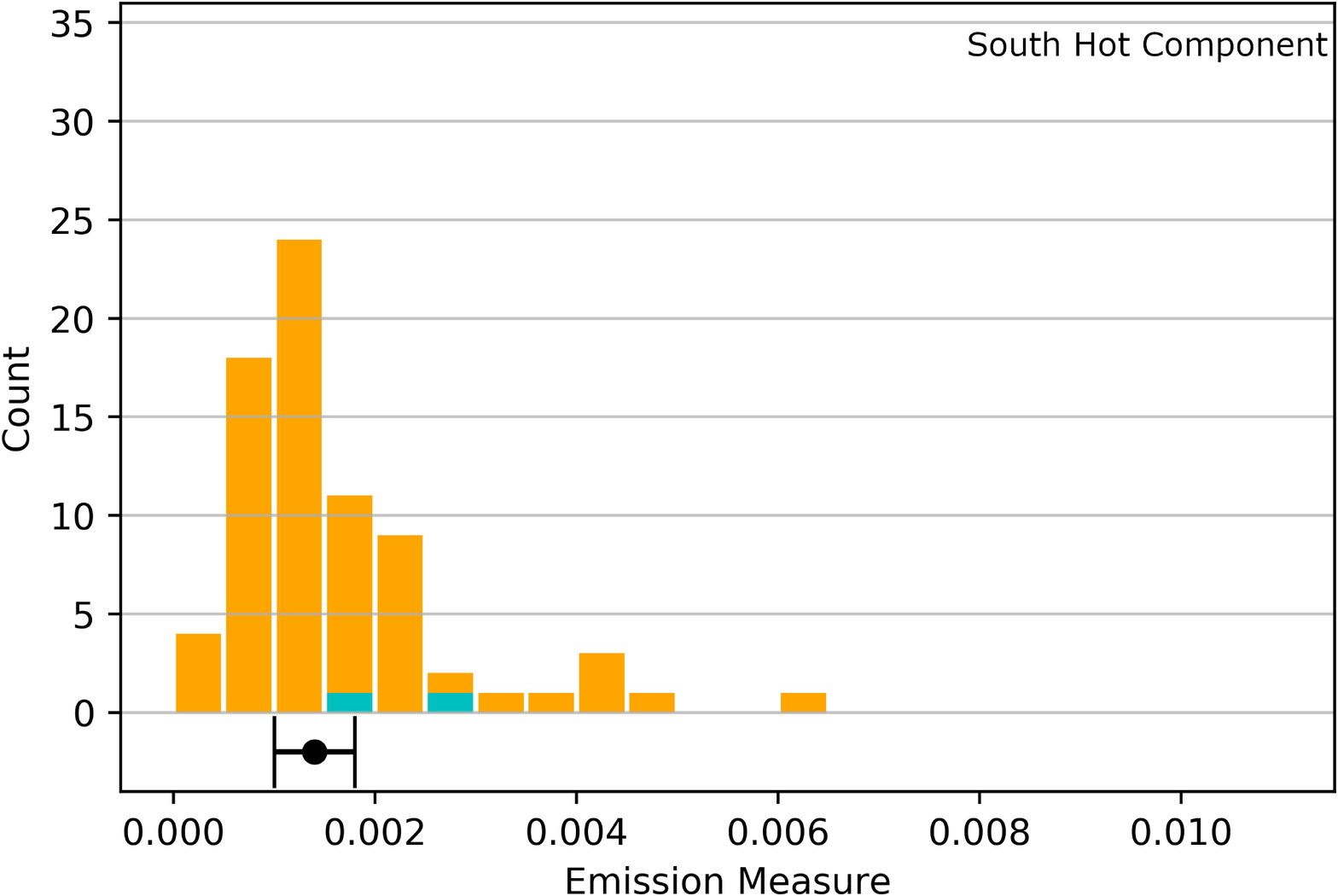} \\
\end{tabular}
\caption{Histograms of emission measure ($\rm cm^{-6}\,pc$) for the northern and southern CGM for both the warm and hot components. The colors follow Figure 3. Near the bottom of each histogram is the median value for the data, with an uncertainty that is equivalent to the average uncertainty (90\% confidence interval). This uncertainty serves to compare the typical uncertainty in the individual measurements to the distribution of measured values.}
\end{figure*}

\begin{figure*}[htb!]
\centering
\begin{tabular}{cc}
  \includegraphics[width=0.49\textwidth]{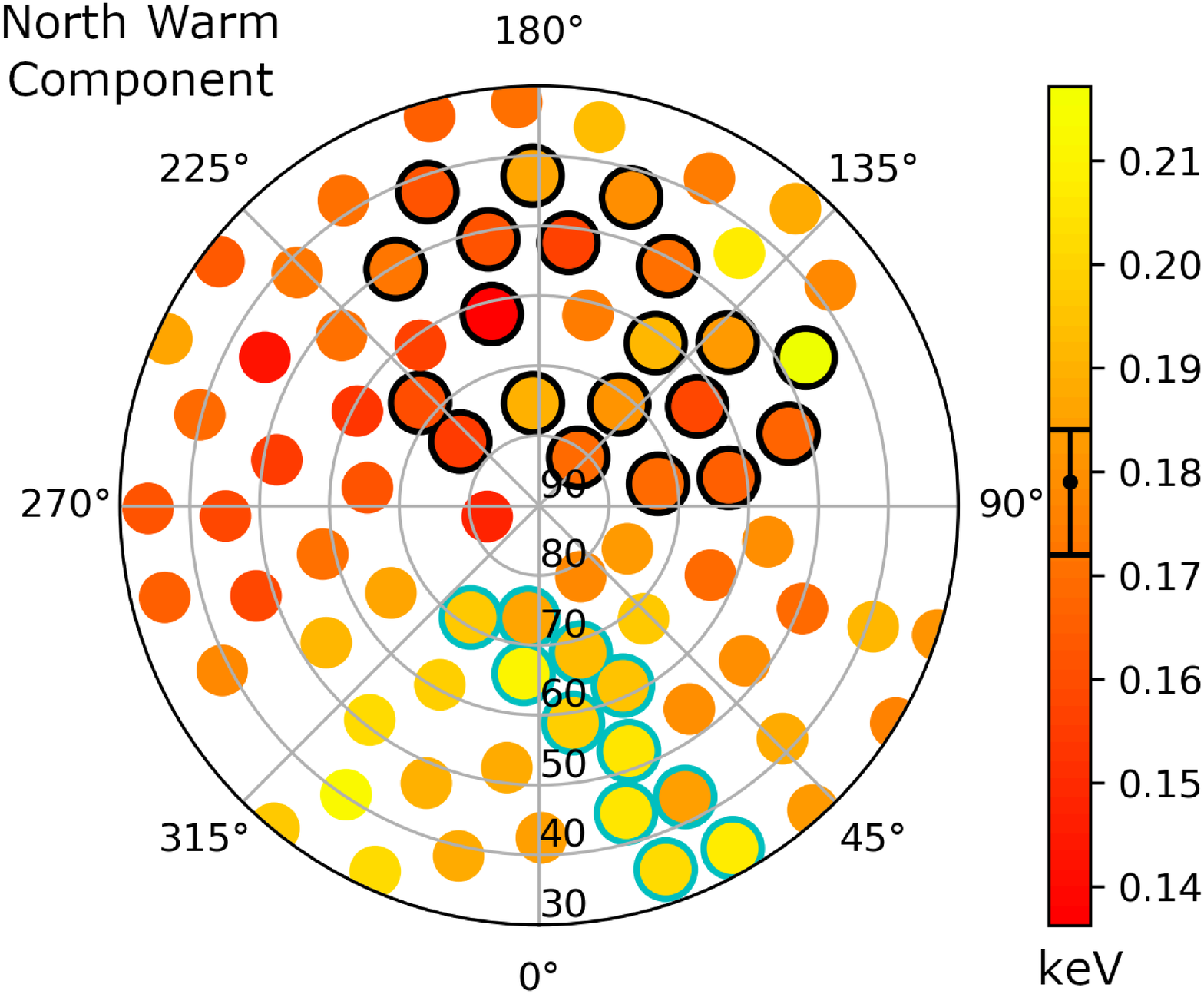} & \includegraphics[width=0.49\textwidth]{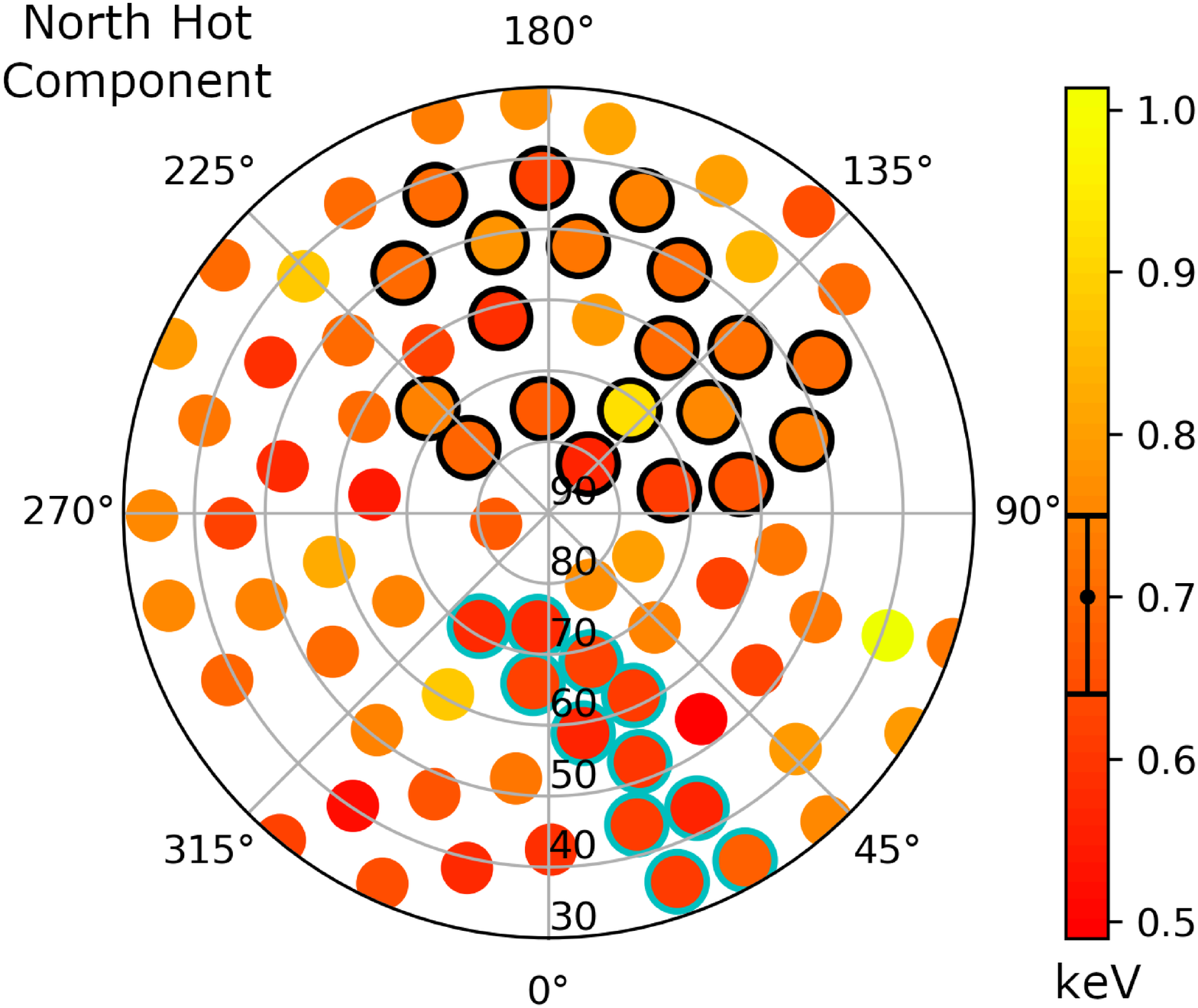} \\
 \includegraphics[width=0.49\textwidth]{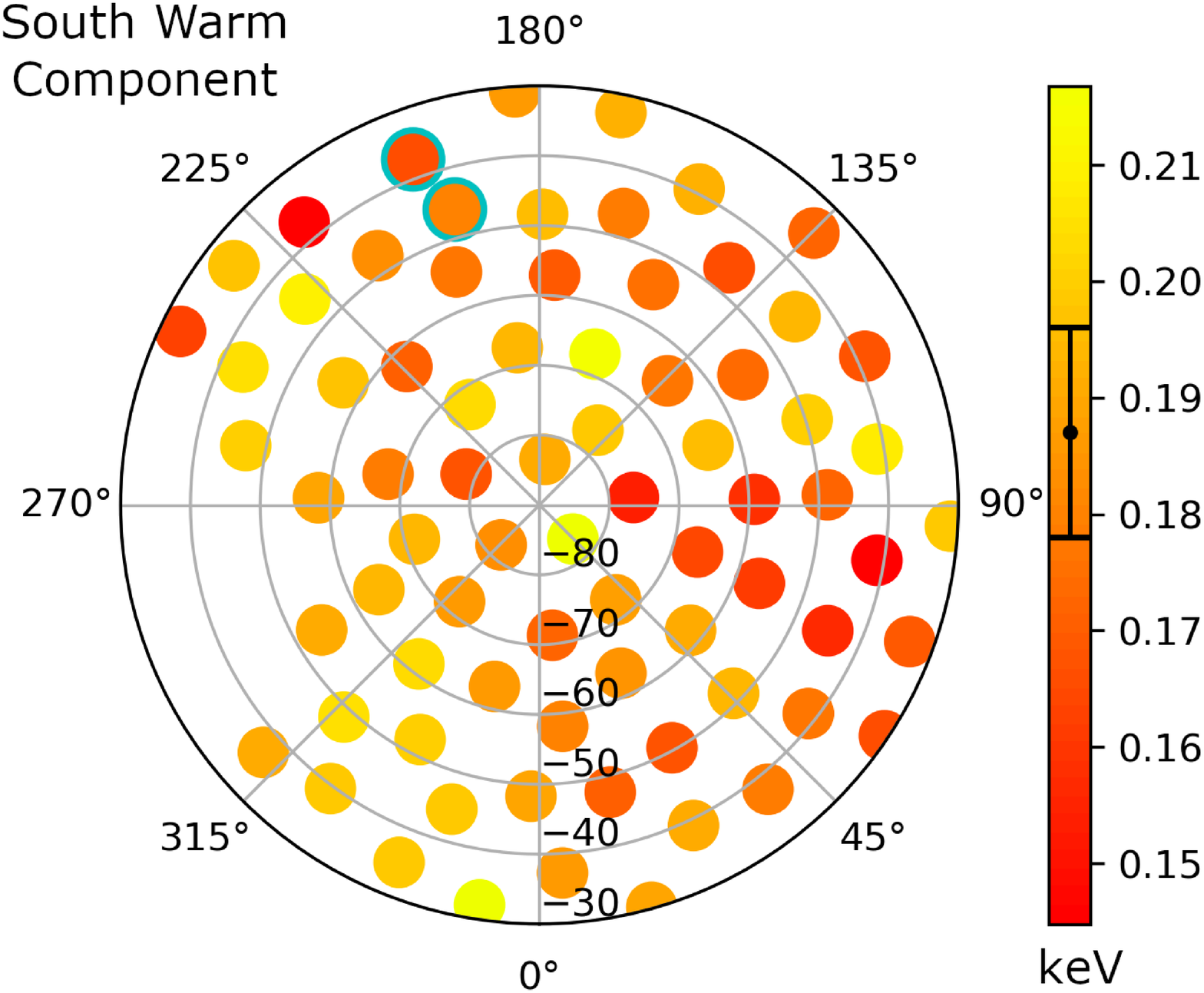} & \includegraphics[width=0.49\textwidth]{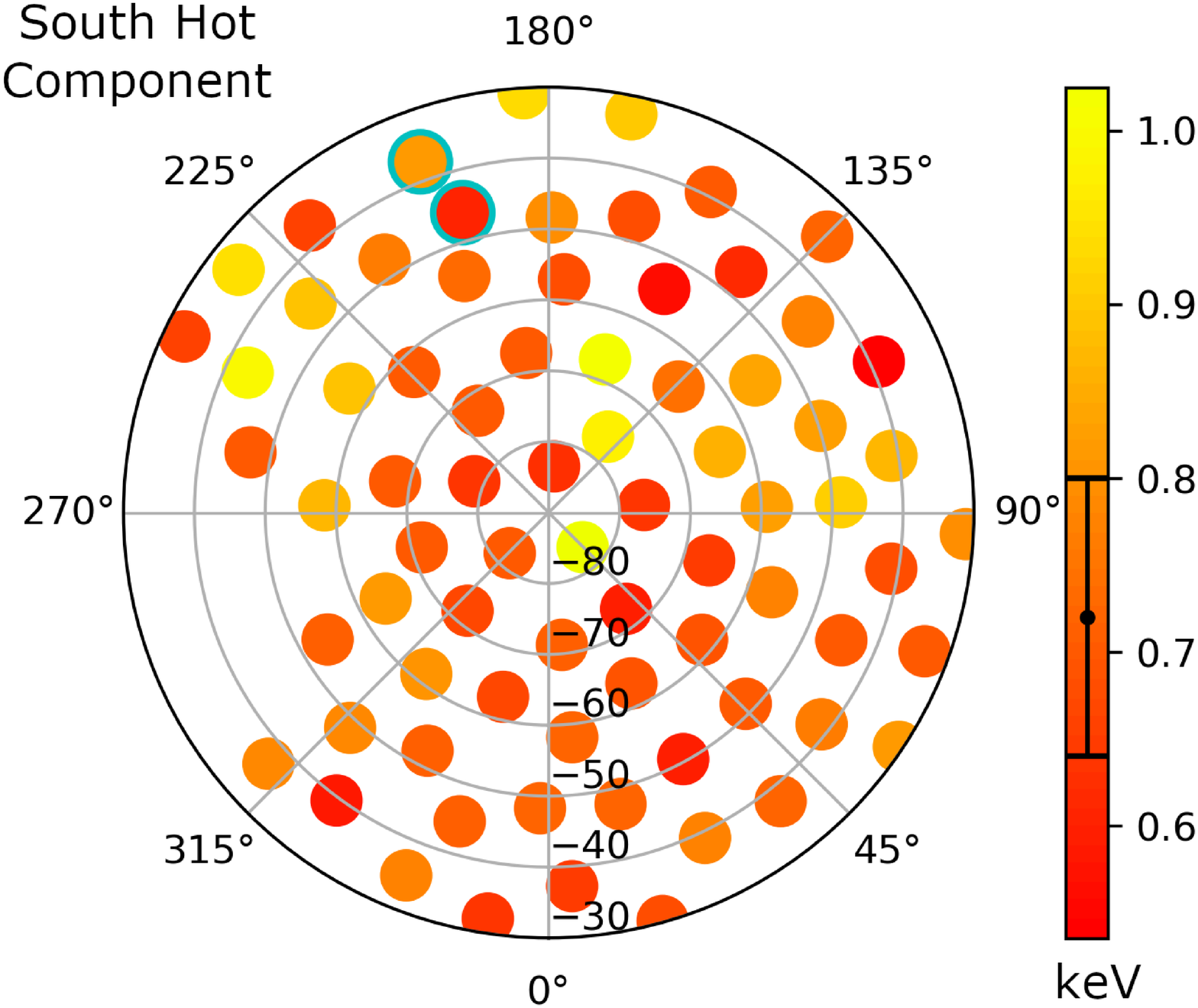} \\
\end{tabular}
\caption{Polar plots for the northern and southern CGM temperatures. Northern CGM figures mark the fields used in the stacked spectra with black rings. The NPS and Eridanus Enhancement are marked with cyan rings. The median temperature from Figure 3 is marked on each color bar, with an uncertainty that is equivalent to the average uncertainty (90\% confidence interval).}
\end{figure*}

\begin{figure*}[htb!]
\centering
\begin{tabular}{cc}
  \includegraphics[width=0.49\textwidth]{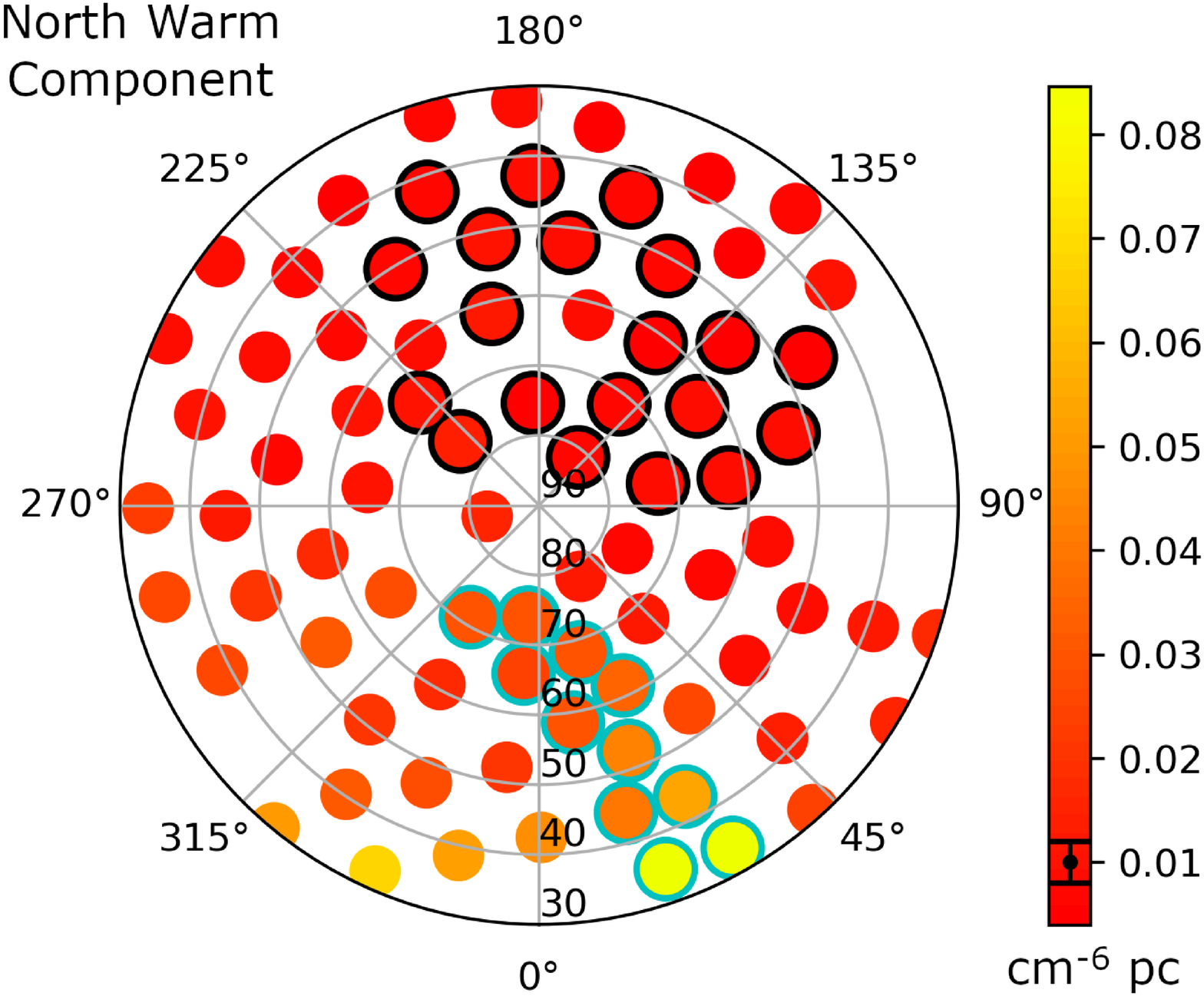} & \includegraphics[width=0.49\textwidth]{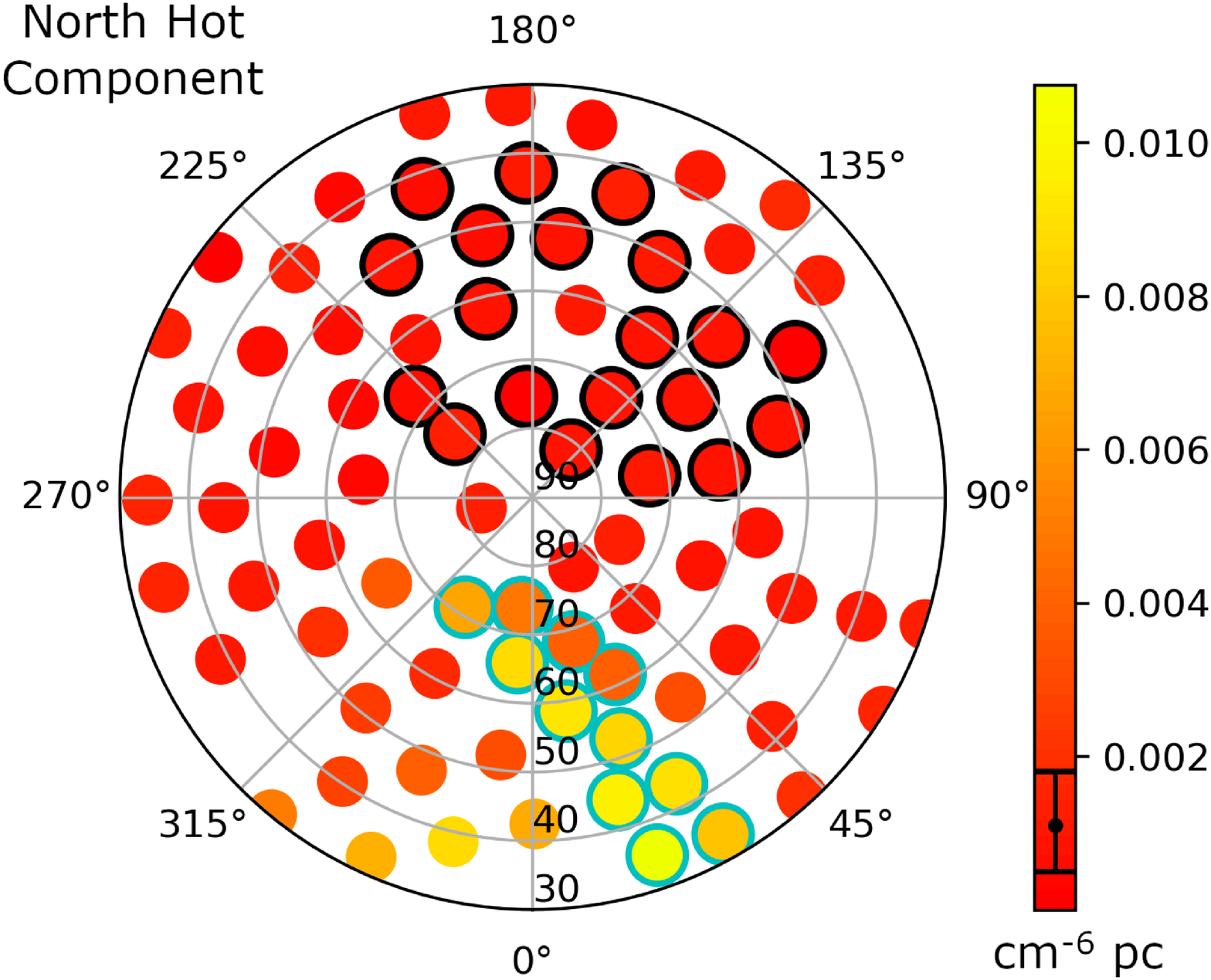} \\
 \includegraphics[width=0.49\textwidth]{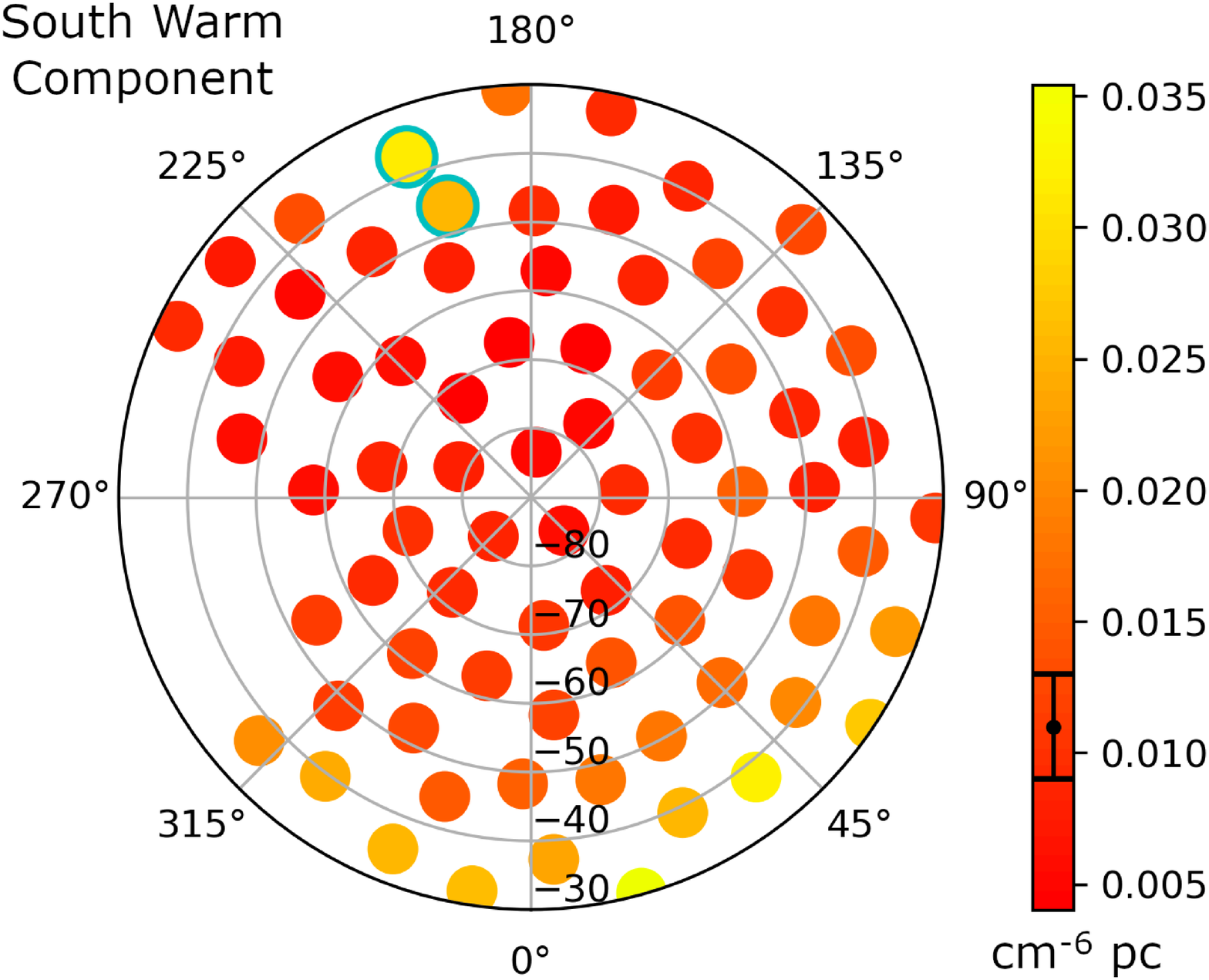} & \includegraphics[width=0.49\textwidth]{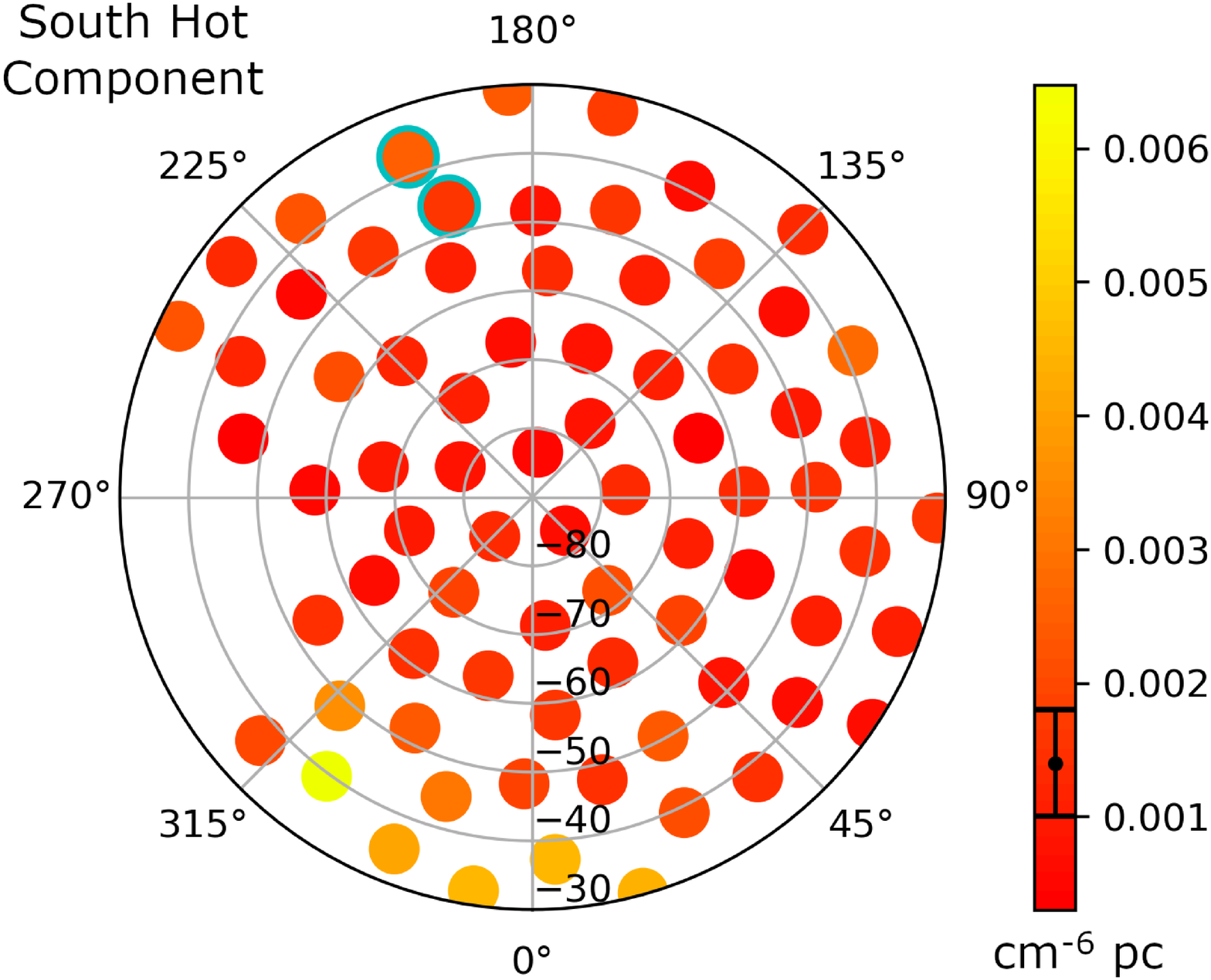} \\
\end{tabular}
\caption{Polar plots for the northern and Southern emission measure. Northern CGM figures mark the fields used in the stacked spectra with black rings. The NPS and Eridanus Enhancement are marked with cyan rings. The median emission measure from Figure 4 is marked on each color bar, with an uncertainty that is equivalent to the average uncertainty (90\% confidence interval).}
\end{figure*}

\section{Analysis} \label{sec:dis}

Fields were initially fit with a single CGM APEC. In many cases, these fits exhibited features in their residuals, with peaks and valleys at the lower energies where the CGM APEC contribution is the strongest, and an unfit bump around 0.8-1.0 keV. However, these initial fit results did reveal a region of the northern CGM with a relatively uniform emission measure and temperature. This motivated stacking these select fields to take a deeper look at the spectral structure of the Galactic CGM and refine the spectral model used in individual fields. Studying the stacked spectra revealed multiple changes in the spectral model, which will be discussed in detail in Section 5, but the biggest takeaway was splitting the CGM APEC into two, with a warm ($\rm \sim 0.18$ keV) and hot ($\rm \sim 0.7$ keV) component. The two CGM APEC stacked spectra model has a $\rm \chi^2$/DoF of 1069/973 (reduced $\rm \chi^2$ = 1.10) improving from a single APEC model fit of 1189/975 (reduced $\rm \chi^2$ = 1.22). These results informed a second round of fitting for the individual fields following the revised fitting strategy used for the stacked spectra.

The final fitting for the individual fields was performed primarily using an automated PyXspec pipeline. Fitting was done in two stages in order to minimize the number of fields for which the fitting ended in local minima. For the first fit, the CGM APEC temperatures were fixed to the values found for the stacked spectra. Afterwards, the CGM APEC temperatures were freed and new fits and error estimates were obtained. After the first iteration of fits were completed, the data was analyzed for consistency, and problematic fields were rerun in a variety of ways depending on the problem. 

There were two sporadic problems apparent in the first iteration of fits. The first was a problem that is common with Xspec fits, where the model is stuck in a local minimum. The second problem was degeneracy between CGM and background components reducing the temperatures of both CGM components in an extreme way. The second and third iterations in the PyXspec pipeline corrected these problems, by using steppar to find the global minima and restricting the temperature range on the hot CGM component. For the final iteration, the hot component temperature was fixed to the stacked spectra value of kT = 0.7 keV only for fields where the temperature of the hot component was poorly constrained.

\section{Results} \label{sec:dis}

Histograms of the temperature values for both components can be seen in Figure 3. Fields with a fixed temperature of 0.7 keV on the hot component have not been included in the hot component histograms. Included near the bottom of each histogram is the average temperature with an uncertainty that is equivalent to the average uncertainty in the individual fields. For the northern CGM histograms, the warm component has an average temperature of kT = $\rm 0.179^{+0.005}_{-0.007}$ keV and a median of $\rm 0.179 \pm 0.013$ keV. The northern hot component has an average temperature of kT = $\rm 0.69^{+0.05}_{-0.06}$ keV and a median of $\rm 0.70 \pm 0.08$ keV. The southern CGM warm component has an average temperature of kT = $\rm 0.184 \pm 0.009$ keV and a median of $\rm 0.187 \pm 0.012$ keV. Meanwhile, the southern hot component has an average temperature of kT = $\rm 0.75 \pm 0.08$ keV and a median of $\rm 0.72 \pm 0.07$ keV. All median errors are median absolute deviation.

The relatively wide distribution of temperatures in the warm component, in comparison to the uncertainty, points towards the varied temperatures being real and not measurement issues. This is less true for the hot component, with its wider error range. A close reader might notice that some fields included in the stacked fit can be seen to be outliers on the upper and lower end of the temperature histograms. This is because the temperatures for those individual fields are not particularly well fit and thus have larger than typical error bars. Those outlier temperatures are actually consistent within error with the stacked spectra temperature.

Histograms of the EM values for both components can be seen in Figure 4 and polar projection maps of the northern and southern CGM temperature and EM can be seen in Figures 5 and 6, respectively. Once again, the median parameter value is included in all figures near the bottom, with an uncertainty that is equivalent to the average uncertainty in the individual fields for that parameter. For the northern CGM, the warm component has an average EM of $\rm 0.018 \pm 0.002$ $\rm cm^{-6}\,pc$ and a median of $\rm 0.010 \pm 0.004$ $\rm cm^{-6}\,pc$. The northern hot component has an average EM of $\rm 0.0023^{+0.0007}_{-0.0006}$ $\rm cm^{-6}\,pc$ and a median of $\rm 0.0011 \pm 0.0004$ $\rm cm^{-6}\,pc$. The southern CGM warm component has an average EM of $\rm 0.013 \pm 0.002$ $\rm cm^{-6}\,pc$ and a median of $0.011 \pm 0.003$ $\rm cm^{-6}\,pc$. The southern hot component has an average EM of $\rm 0.0016 \pm 0.0004$ $\rm cm^{-6}\,pc$ and a median of $\rm 0.0014 \pm 0.0005$ $\rm cm^{-6}\,pc$. All median errors are median absolute deviation. The noticeably non-normal distribution of the emission measures makes the averages and medians much more different here than they were for the temperatures. The median values are more appropriate in this case. Note that the northern CGM average and median values do include the NPS fields, which are significant outliers in the histograms. Regardless of the NPS, the wide variation in EM for both the warm and hot components show that the CGM features a clumpy distribution, consistent with the conclusions of \citet{Kaaret2020}. 

The temperatures of the components in the south appear greater than those of the north, and the emission measures of the components in the south appear less than those of the north. However, a two-sample Kolmogorov-Smirnov (KS) test on the temperature of the warm and hot components of the CGM for the north and south returns p-values of 0.19 and 0.25, respectively. Note that fields with the temperature of the hot component fixed to kT = 0.7 keV were removed from the KS test of the hot component. A KS test for differences in the distribution of EM returns p-values of 0.17 and 0.18 for the warm and hot components, respectively. As such, the KS does not provide evidence for a statistically significant difference in temperature or EM between the north and south for either CGM component.

\begin{figure*}[htb!]
\centering
\includegraphics[width=0.75\textwidth]{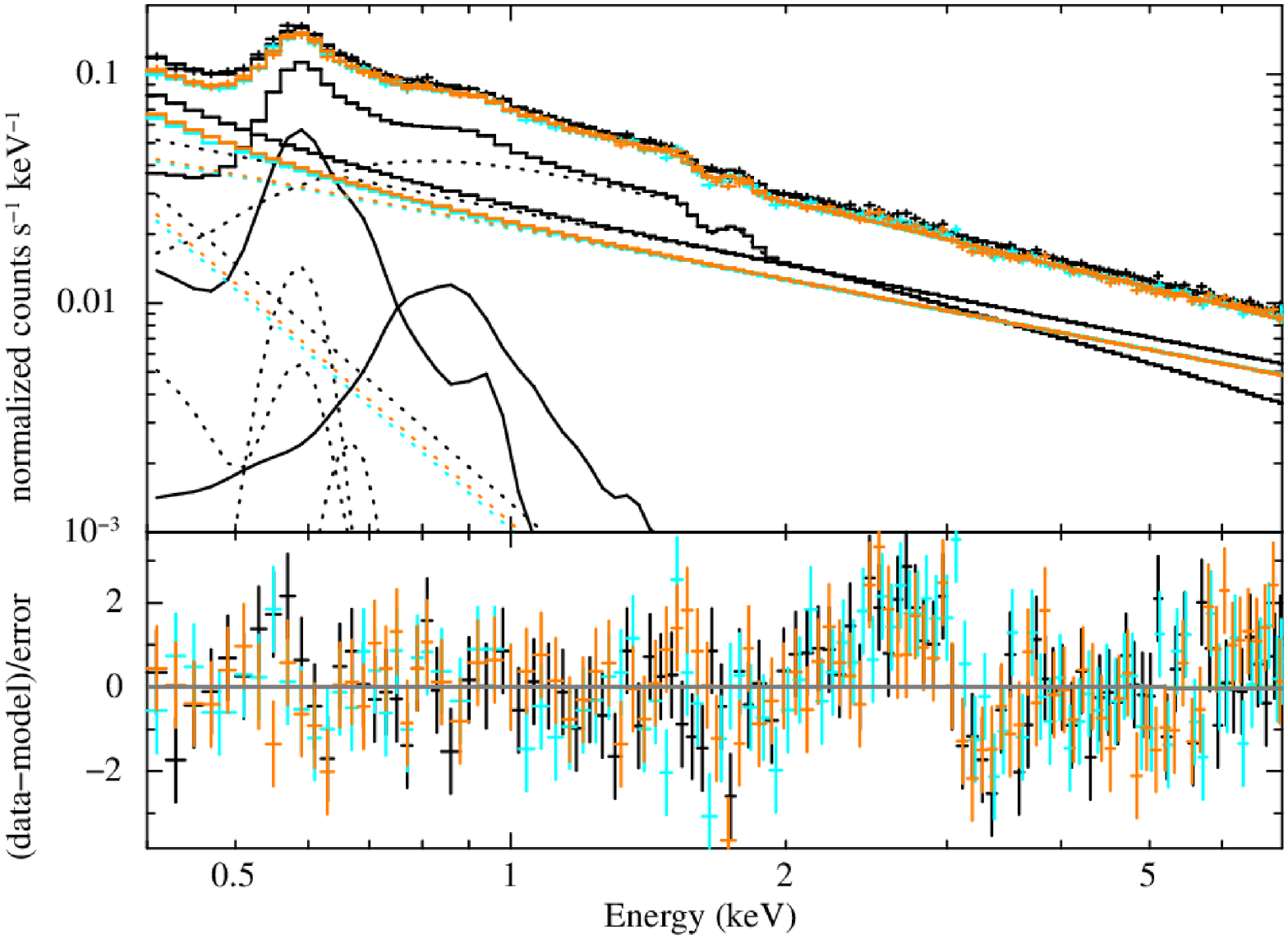}
\caption{Stacked spectra of consistent fields in the northern CGM. Each detector is a different color (DPU 14, 54, and 38 are respectively black, orange, and cyan). The model for the CGM is two APECs, marked in solid black lines. Background and foreground components are marked in dashed lines. Model components that are identical between each detector are only shown once. The feature around 3 keV is due to the data filtering procedure as described in the text.} 
\end{figure*}

\begin{figure*}[htb!]
\centering
\includegraphics[width=0.75\textwidth]{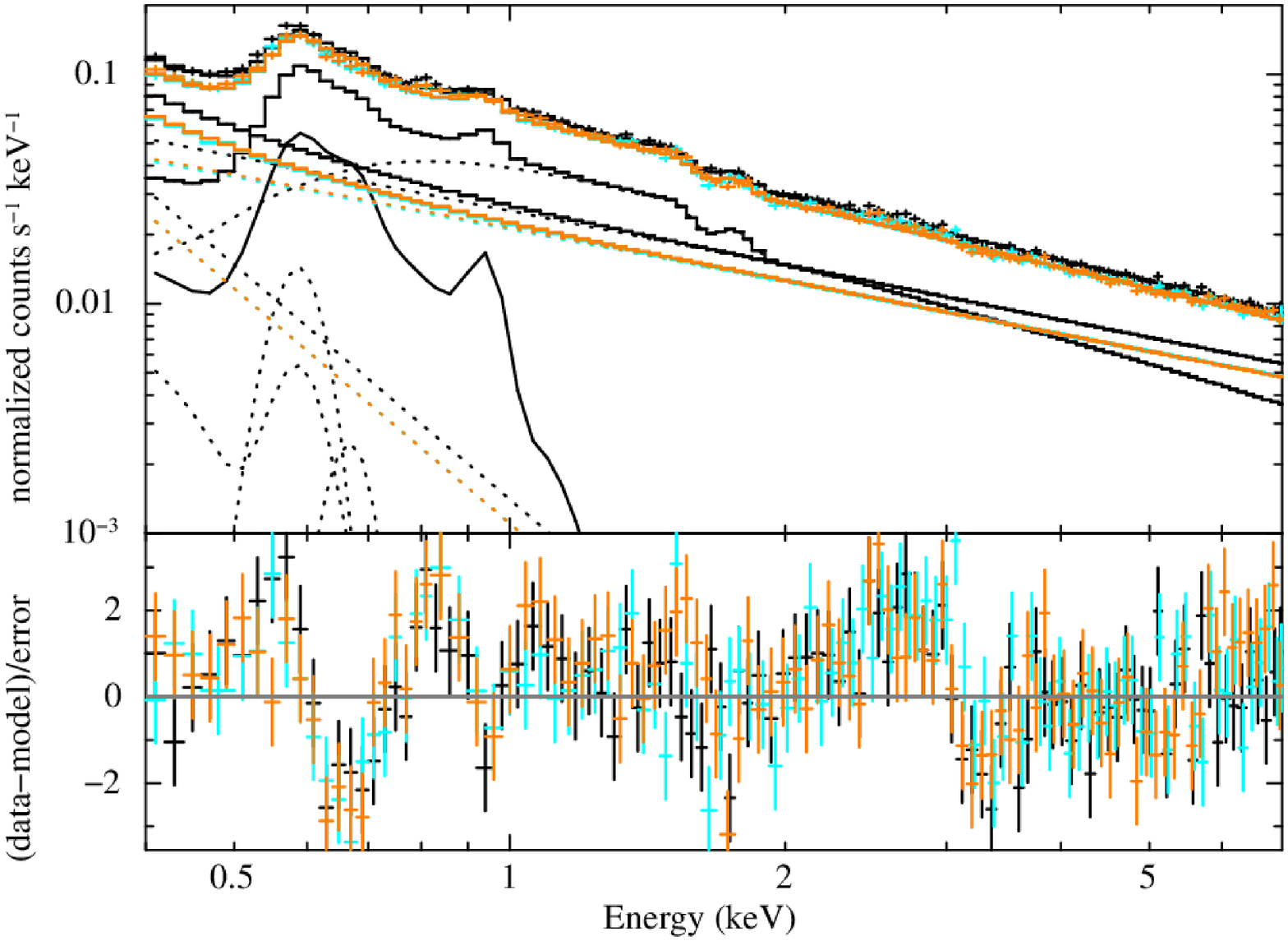}
\caption{Stacked spectra of consistent fields in the northern CGM. Each detector is a different color, following Figure 7. The model for the CGM is a VAPEC with freed neon abundance, marked with a solid black line. Background and foreground components are marked in dashed lines. Model components that are identical between each detector are only shown once. This overall fit is worse than the fit presented in Figure 6. There are significant residuals around 0.5-0.7 keV and a poorly fit section of the high temperature bump from 0.8-0.9 keV. This is the region that corresponds to the neon emission. The additional high temperature APEC fit in Figure 6 fits to this feature much better than the neon enhancement shown here does.} 
\end{figure*}

\section{Stacked Spectra} \label{sec:dis}

The extended region with consistent EM seen in the northern CGM allowed the stacking of multiple fields together to study the CGM spectrum to a deeper degree. The fields that were included in the stack are marked with black rings in Figures 4 and 5. After a good model was found for the stacked spectra, the model and stacked spectra parameters were tested on the individual fields and four fields were identified that were not consistent with the stacked model and parameters. These fields were removed from inclusion in the stack and the revised stacked spectra was refit. 

Out of the four fields removed from the stacked spectra, three had apparent data quality issues that affected the fit. Those fields preferred non-physical fits with abnormal temperatures. Note that these selections were performed before the foreground/background analysis and response parameters were revised, as those were motivated by the stacked spectra. These fields were recovered with physically reasonable fits in later fitting iterations for the individual field fits. The fourth field (HS0199) preferred to fit with reasonable, but different temperatures than those found in the stacked spectra.

The model used was ultimately the same as the individual field fits in terms of foreground and background model components (with an unabsorbed LHB apec, absorbed CXB power-law, and two particle background power-laws, plus two SWCX Gaussians). The fixed parameters in the model fit were averaged from the values for each field, weighted by the total exposure for all three detectors for the field. The particle background power-laws were fit in the same way as they were for the individual fields, as analysis showed that the sum of the individual paired power-laws from each field could be closely fit by a simple pair of power-laws.

The stacked spectra represent an exposure of 788 ks, 829 ks, and 824 ks for DPU 14, 54, and 38, respectively, for a total of 2.44 Ms. The high statistical quality of this exposure makes the analysis sensitive to subtle effects in the instrument response, that were not recognized in \citet{Kaaret2020}. In advance of our CGM analysis, an analysis was carried out of the supernova remnants Cassiopeia A. Cas A is very bright in the soft X-ray band and was a HaloSat calibration target observed repeatedly over the course of the mission. Analysis of the Cas A spectra revealed an offset in the channel to energy conversion (or `gain'). Further analysis of the Crab also revealed the need for the addition of a silicon absorption edge to the response matrix. The Crab is another HaloSat target with a substantial depth of data and is known for having a featureless X-ray spectra, and thus any missing edge component will readily stand out. For DPUs 14, 54, and 38 gain offsets of 0.0232, 0.0240, and 0.0239 were found. The silicon edge has a threshold energy of 1.839 keV and an absorption depth of -0.170. Further details can be found in Appendices A and B. The revisions to the gain and the response were applied to both the stacked spectra and the spectra for the individual fields.

The best fit for the stacked spectra was the two APEC model, with a warm and hot component for the CGM. The stacked spectra can be seen in Figure 7. The stacked spectra residuals are relatively smooth, with the exception of the edge around 3 keV due to the 3-7 keV filtering procedure (if one selects all spectra with a band value less (greater) than a given value, then the resulting spectrum will have a decrement (increment) in exactly that band compared to the adjacent energies, just due to the independence of the uncertainties between individual bins). This artifact is only noticeable in this stack fit due to the large amount of time included in the stack. The fit parameters for the two APEC model can be found in Table 1. The stacked spectra 90\% confidence errors were not generated using MCMC. The final fit has a $\rm \chi^2$/DoF of 1069/973, for a reduced $\rm \chi^2$ of 1.10. Using a one APEC model, with freed background parameters, changes the fit statistic to 1189/975 = 1.22. 

\citet{Gupta2020} provides an alternative explanation for the observed hot component: an enhancement of neon relative to the rest of the CGM abundances. To test this possibility, the pair of CGM APECs were replaced by a single variable abundance APEC (VAPEC) model (XSPEC model {\tt vapec}), allowing for the abundance of neon to be a free parameter in the fit. Note that the absorption abundance in the model was not changed, as the neon enhancement was treated as a feature of the CGM and not the interstellar medium. For consistency, the background parameters were fixed to the values from the two APEC fit. The neon abundance fits to a value of $\rm 0.71^{+0.10}_{-0.09}$ solar. The double APEC model is a significantly better fit ($\rm \chi^2$/DoF = 1069/973) when compared to the VAPEC model ($\rm \chi^2$/DoF = 1241/984). There are also noticeable features in the residuals, see Figure 8. The enhanced neon peak fits the emission from 0.9-1.0 keV nicely, but fails to account for the enhanced emission from 0.8-0.9 keV, whereas the full 0.8-1.0 keV energy range is easily matched by the peak of the hot APEC. Even if the background parameters are freed and allowed to fit, the fit statistic only improves to 1150/974, and the residual problems remain. Ultimately, the comparison between the VAPEC and additional APEC is independent of the background fitting, as there is no way for power-laws to generate the peaked features seen in the residuals. It is for this reason that we consider the double APEC model to be the preferred model for the CGM.

Another open question relating to abundance is the actual metallicity of the CGM. \citet{Kaaret2020} treated the metallicity of the CGM as 0.3 solar, as does this paper, but \citet{Kaaret2020} and other studies (such as \citet{Mitsuishi2012, Yoshino2009}) have also opted for using or checking values as high as 1.0 solar. We fixed the background to the original fit values before adjusting the abundances. The change in abundance results in no significant change to the fit temperatures, as it is essentially degenerate with the APEC normalization. The quality of the fit worsens with the $\rm \chi^2$/DoF increasing from 1069/983 for the 0.3 solar fit to 1082/983 for solar abundance, due to minor shifts in the residuals at the lowest energies.

We also tested freeing the abundances in the two APEC fit, which resulted in the hot CGM component being unable to fit an abundance, while the warm component fit to $\rm 0.27^{+0.18}_{-0.08}$, a value consistent with the fixed value. The fit parameters are essentially the same, and the fit statistic changed to a $\rm \chi^2$/DoF of 1068/982, essentially the same value. Given the ambiguities of fitting an abundance in this way, we do not prefer a model where the abundance is free to fit.

\begin{deluxetable}{ccc}
\tablenum{1}
\tablecaption{Stacked Fit Parameters\label{tab:MP}}
\tablewidth{0pt}
\tablehead{
\colhead{model} & \colhead{parameter} & \colhead{value}\\
}
\startdata
Warm CGM APEC & kT (keV) & $\rm 0.166 \pm 0.005$\\
  & EM ($\rm cm^{-6}\,pc$) & $\rm 0.0129^{+0.0009}_{-0.0008}$\\
  \\
Hot CGM APEC    & kT (keV) & $0.69^{+0.04}_{-0.05}$\\
  & EM ($\rm cm^{-6}\,pc$) & $\rm 0.0013 \pm 0.0002$\\
  \\
\hline
Power law 1 (DPU 14)& photon index & $\rm 0.79 \pm 0.03$\\  
 & normalization & $0.0254^{+0.0009}_{-0.0013}$\\
\\
Power law 1 (DPU 54)& photon index & $\rm 0.77 \pm 0.03$\\  
 & normalization & $0.0214^{+0.0008}_{-0.0011}$\\
\\
Power law 1 (DPU 38)& photon index & $\rm 0.76 \pm 0.03$\\  
 & normalization & $0.0211^{+0.0008}_{-0.0010}$\\
\\
Power law 2 & photon index & $3.4^{+1.0}_{-0.8}$\\  
 & normalization (DPU 14) & $0.0014^{+0.0018}_{-0.0008}$\\
  & normalization (DPU 54) & $0.0011^{+0.0015}_{-0.0007}$\\
   & normalization (DPU 38) & $0.0011^{+0.0013}_{-0.0006}$\\
\\
\hline
Fit (DPU 14) & $\rm \chi^2$ & 359\\
Fit (DPU 54) & $\rm \chi^2$ & 369\\
Fit (DPU 38) & $\rm \chi^2$ & 341\\
Fit total & $\rm \chi^2/DoF$ & 1069/973\\
\\
\enddata
\tablecomments{Row 1 is the name of the model component. Row 2 is the name of the parameters for the component. Row 3 is the value for the listed parameter. The top section is the astrophysical components while the middle section is the instrumental components, and the bottom section is the fit statistics. Errors are the 90\% confidence interval.}
\end{deluxetable}

\section{Discussion} \label{sec:res}

\begin{figure*}[htb!]
\centering
\begin{tabular}{cc}
\includegraphics[width=0.49\textwidth]{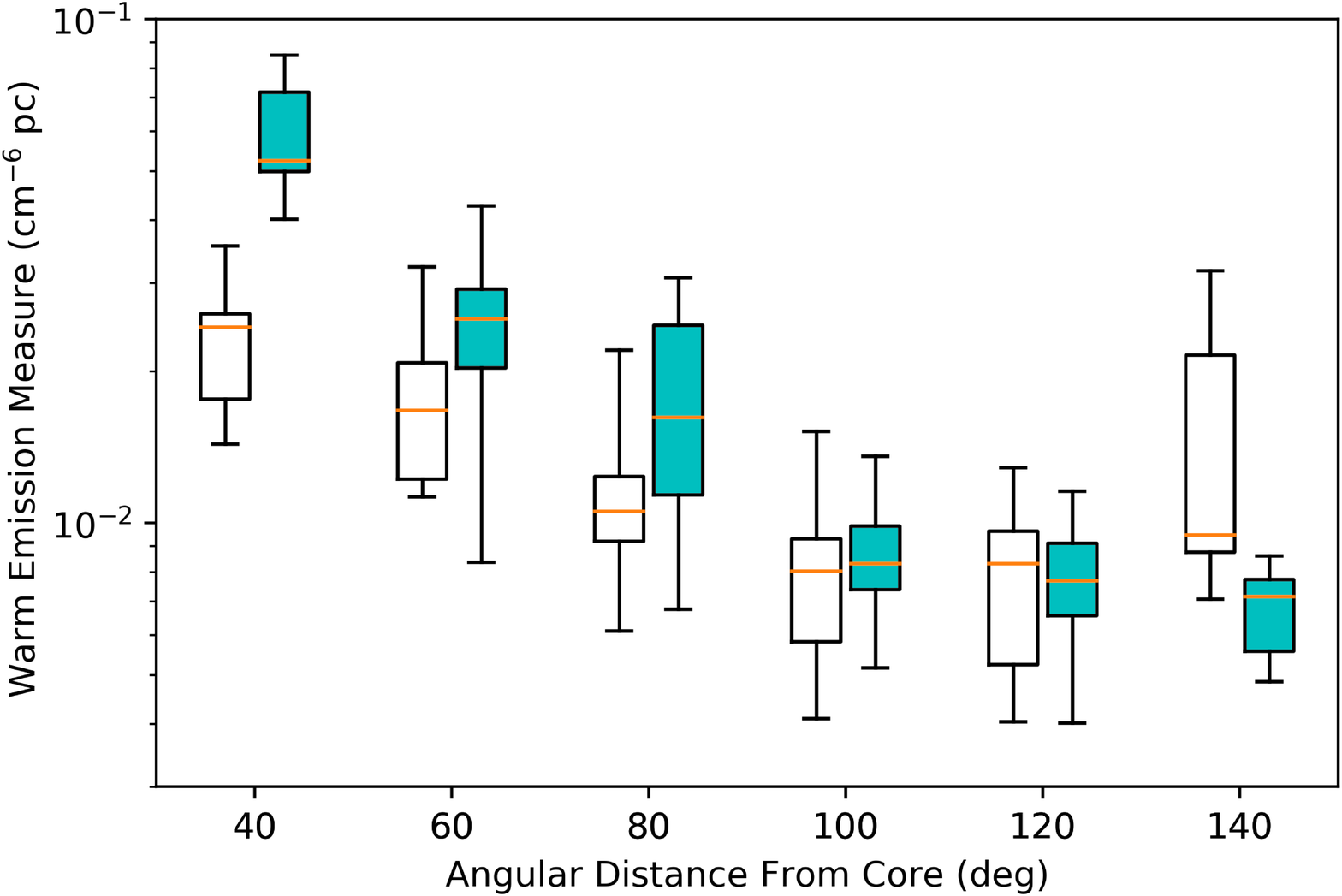} & \includegraphics[width=0.49\textwidth]{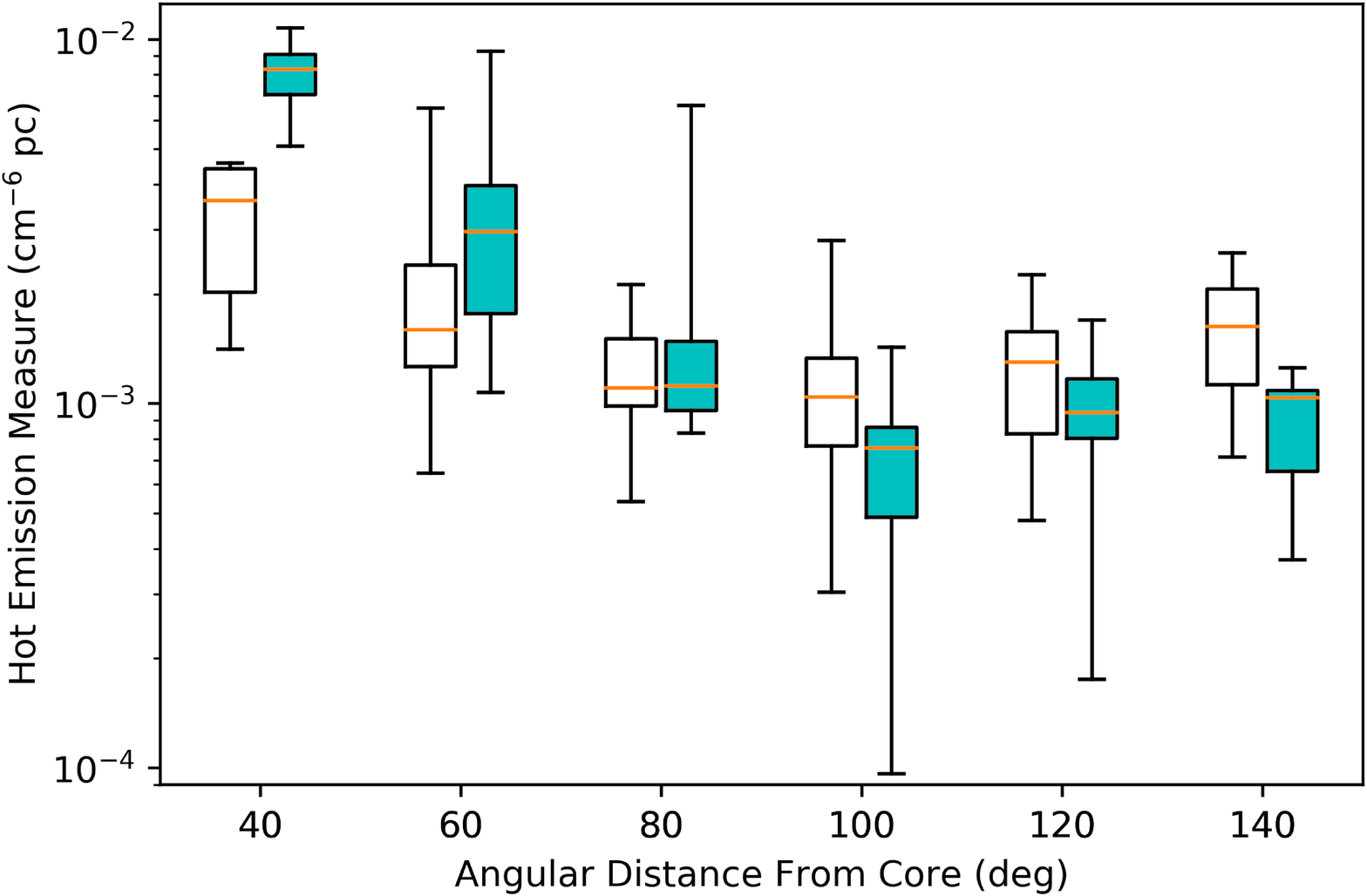} \\
\end{tabular}
\caption{Box-whisker plots of warm (left) and hot (right) emission measure versus angular distance from the Galactic center. Bin width is 20 degrees. For each bin, the northern and southern CGM are plotted on each side of the bin center-point. Boxes for the northern CGM are filled with cyan while the southern CGM boxes are unfilled. The box represents the two inner quartiles for each bin and the orange line in each box is the median. The outer whiskers on each box are the two outer quartiles of the binned data.}
\end{figure*}


\begin{figure*}
\centering
\includegraphics[width=1\textwidth]{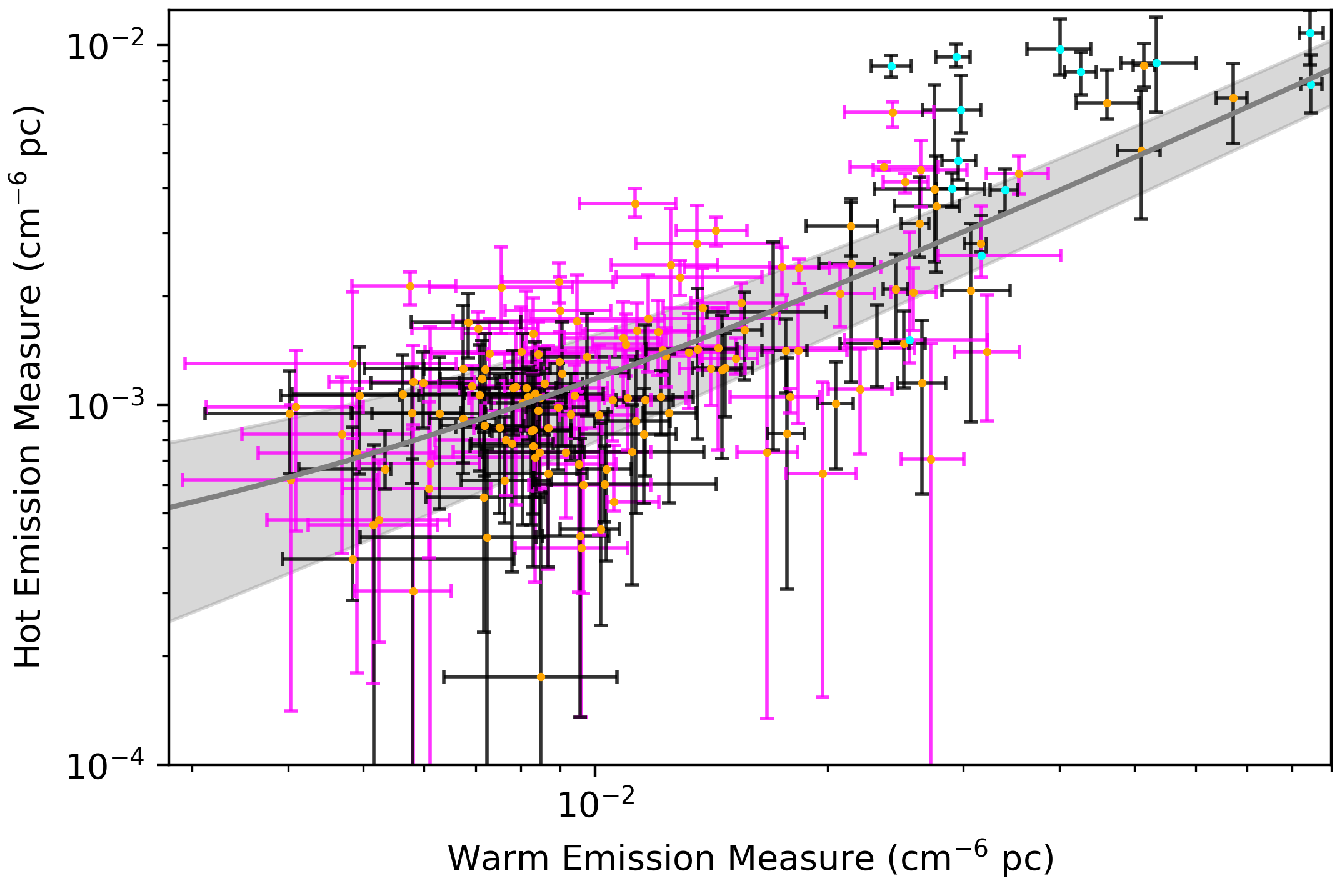}
\caption{Log scale plot comparing the warm component emission measure to that of the hot component. Northern CGM fields are marked with black error bars, while the southern CGM fields are marked with magenta error bars. NPS and Eridanus Enhancement fields are marked with cyan center points, while other fields are marked in orange. The gray line is a regression line for the data based on the Bayesian method from \citet{Kelly2007}, which utilizes the error on both components. The shaded grey region is the 90\% confidence region for the regression. The regression line does not include the NPS or Eridanus Enhancement, and also does not use the 4 additional outlying fields in the upper right (which are near the Galactic bulge). The linear regression has a slope of $\rm 0.092 \pm 0.017$ and an intercept of $\rm 0.00026 \pm 0.00022$ (90\% confidence interval), calculated using the linmix python module from https://github.com/jmeyers314/linmix. The slope being close to 0.1 reflects the typical ten-fold difference between the two component's emission measures.}
\end{figure*}

Using X-ray observations from HaloSat, we have detected a secondary hotter component in the spectrum of the CGM across a large fraction of the northern and southern CGM. Evidence for the hot component comes from the stacked spectra of northern fields and from the spectra of the individual fields. The stacked spectra represent 25\% of our fields in the north, all of which are far away from the Galactic bulge. The detection of the hot component in the stacked spectra is robust, and has a temperature of kT = $0.69^{+0.04}_{-0.05}$ keV. The detection of the hot component in the individual fields is also robust, with 85\% of the fields (133 out of 156) having a detection significant of $\rm 3\sigma$ or more. This significance was evaluated by examining the ratio of the hot component EM to its lower error bar. The hot component temperature is roughly consistent within errors across the CGM, with an average temperature of kT = $\rm 0.69^{+0.05}_{-0.06}$ keV for the northern CGM, and kT = $\rm 0.75 \pm 0.08$ keV for the south. Note that these errors are the average of the errors for the individual fields. The north, south, and stacked spectra temperatures are all consistent with each other.

Previously this hotter component had only been seen for discrete sight-lines, \citep{Gupta2020,Das2019a,Das2019c}, many of which are near the Galactic core. Our typical temperature ranges (see Figure 3) for both the warm (kT $\rm \sim 0.15$ to $\rm \sim 0.21$ keV, $\rm 0.166 \pm 0.005$ keV for the stacked fit) and hot components (kT $\rm \sim 0.5$ to $\rm \sim 1.0$ keV) are consistent with the results from \citet{Gupta2020} of kT = $\rm 0.176 \pm 0.008$ keV and $\rm \sim 0.65$ to $\rm \sim 0.90$ keV, which are also based on emission measurements. They are also similar to the emission-based temperature range from \citet{Das2019c} of kT $\rm \sim 0.15$ to $\rm \sim 0.23$ keV and $\rm \sim 0.4$ to $\rm \sim 0.7$ keV. The absorption-based hot component temperature ($\rm \sim 0.99$ keV) of \citet{Das2019b, Das2019c} is consistent with our hot component temperature as well. Their absorption-based warm component has a center-point temperature of 0.111 keV, apparently inconsistent with our work, although the upper error range stretches up to 0.176 keV. \citet{Das2019c} consider the absorption-derived warm component to be a third temperature component in the CGM. 

\citet{Nakashima2018} fit X-ray observations with a single temperature model for the CGM and found that a small selection of observed fields fit to higher CGM temperatures around 0.4 or 0.7 keV. That work included a component at kT = 0.1 keV with free normalization for local SWCX and LHB emission that likely contributed to that fit result, partially masking the warm CGM component. This would have resulted in some fields fitting the CGM partially or totally to the hot component instead of the stronger warm component. This is a result that we observed in our early fitting, prior to making improvements to our background models and model parameters. For those early fits, a small number of fields ended up fitting with a low temperature APEC similar to the LHB component and a higher temperature component at roughly 0.4 (combining part of the warm component emission with the hot component) or 0.7 keV (just fitting the hot component peak). The higher kT fields from \citet{Nakashima2018} appear to be a random subset and exhibit no notable properties in our data.

The X-ray Quantum Calorimeter (XQC) is an instrument flown on a sounding rocket that provided high-resolution X-ray data for the diffuse X-ray background, including the CGM \citep{McCammon2002}. It has been flown multiple times, including observations during 1999 and 2008 targeting l = $\rm 90^{\circ}$ and b = $\rm 60^{\circ}$, which aligns with our data set. The XQC field-of-view is 1 steradian, so it includes quite a few HaloSat fields. However, the XQC field does include a few HaloSat fields near the ecliptic pole that are removed by our sun angle cut in order to reduce the contributions from SWCX. \citet{Wulf2019} reanalyzes the XQC fields and report line strengths for the strong \ion{O}{vii} and \ion{O}{viii} lines that contribute to the LHB, SWCX, and CGM. We can compare line strengths for the CGM derived from HaloSat to those from XQC to check for consistency. \citet{Wulf2019} estimates that 37\% of the oxygen line contribution is from the CGM. This makes the estimated 1999 CGM line strengths for \ion{O}{vii} and \ion{O}{viii} $\rm 1.75^{+0.56}_{-0.52}$ line units and $\rm 0.55^{+0.29}_{-0.23}$ line units, respectively. The 2008 observation (with the individual \ion{O}{vii} lines summed) is similarly $\rm 1.57^{+1.65}_{-0.70}$ line units for \ion{O}{vii} and $\rm 0.34^{+0.18}_{-0.14}$ line units for \ion{O}{viii}. The HaloSat \ion{O}{vii} and \ion{O}{viii} line strengths (averaged across all of our included fields within the XQC field-of-view) are $\rm 2.14 \pm 0.06$ line units and $\rm 0.64 \pm 0.02$ line units respectively. These lines are consistent with the values reported in \citet{Wulf2019} for 1999 and 2008, with the exception of the 2008 \ion{O}{viii} lines. The 2008 \ion{O}{viii} lines are still close and the difference could come down to the estimate for the CGM contribution.

It is also interesting to compare higher energy neon and iron lines from our CGM data to the \citet{Wulf2019} results as a check on the existence of our hot component. While there are hints of neon and iron lines in the XQC data, the presence of those lines in the XQC data is inconclusive. Line strengths for \ion{Ne}{ix}, \ion{Ne}{x}, and \ion{Fe}{xvii} were calculated for the HaloSat models and compared to the 1999 and 2008 XQC observations. In both cases, all three lines were found to be below the noise level of those observations. As such, the relatively short duration XQC observations are consistent with our relatively weak hot component.

\citet{Wulf2019} also analyzed observations taken of a field positioned along the plane of the Galaxy. These observations exhibited a hot component somewhat similar to what we see for our CGM fields, but \citet{Wulf2019} interprets this high temperature contribution as stemming from M stars in the Galactic plane. They include no such enhancement in the XQC field that overlaps with the HaloSat fields studied in this paper. \citet{Masui2009} also studies the X-ray contributions of M stars along the Galactic plane, modeling the excess emission as an APEC, somewhat similar to what we observe for the CGM. However, that work notes that the M star contribution drops off steeply, diminishing exponentially over just a few degrees of latitude. Previous investigation of stellar contributions to the HaloSat spectra \citep{Ringuette2021} has not supported a significant contribution from stars. The fields studied in this paper are even higher latitude than the fields from \citet{Ringuette2021}. Due to the selection of only high latitude fields for our CGM analysis, it is unlikely that M stars are a significant contributor to our hot component.


In comparison to the previous CGM work using HaloSat from \citet{Kaaret2020}, we find a cooler warm component than that work's value of $\rm \sim 0.22$ keV. This is in large part due to the splitting of the single CGM APEC into two, which consistently reduces the temperature of the warm component. The HaloSat based results from \citet{Ringuette2021} are somewhat inconsistent, although the fields in that paper are closer to the Galactic plane than the fields studied in this paper (b = $\rm -16^\circ$, $\rm -24^\circ$). They reported CGM temperatures of kT = $\rm 0.26^{+0.03}_{-0.02}$, $\rm 0.262^{+0.016}_{-0.014}$ keV and kT = $\rm 1.01^{+0.13}_{-0.11}$, $\rm 1.03^{+0.23}_{-0.19}$ keV. The warm component is inconsistent with the range of temperatures seen for the warm component in this paper, while the hotter component is not consistent with the stacked spectra, but is within the range of temperatures seen in individual fields. It is important to note that \citet{Ringuette2021} does not include the changes made to the instrument response outlined in this paper's appendices, and models some of the foreground and background components differently due to very different scientific objectives.

The CGM temperatures we have found can also be compared to other galaxies. NGC 891 is an edge-on galaxy sometimes regarded as a Milky Way analog, and has been fit with a two temperature model in \citet{Hodges2018}. In that paper, a warm component at kT = $\rm 0.199 \pm 0.008$ keV is found, along with a hot component at kT = $\rm 0.71 \pm 0.04$ keV. This hot component temperature is quite consistent with the hot component detected in this paper, with an average value of 0.69-0.75 keV. The warm component temperature is inconsistent with the stacked spectra, but as seen in Figure 3 the stacked spectra consists primarily of fields on the cooler side of the temperature range, and the warm component of NGC 891 is consistent with the wider range of temperature in the individual fields. \citet{Hodges2018} find that the hotter component is more concentrated above star-forming regions in the galactic disk while the warm component is more widespread. The hot component is far enough above the disk that it cannot be explained by contributions from M stars. Our temperature range is also consistent with the temperatures of the late-type spiral galaxies M51 and M83 of $\rm \sim 0.2$ and $\rm \sim 0.65$ keV from \citet{Owen2009}.

On the other hand, the results of \citet{Tullman2006} (9 galaxies ranging between quiescent and starbursting, with 6 having measured halo temperatures) and \citet{Strickland2004} (10 galaxies, 7 of which are starburst) appear mostly inconsistent with our results. Those studies found lower temperatures for both components at kT = $\rm \sim 0.06$ to $\rm \sim 0.17$ keV and kT = $\rm \sim 0.19$ to $\rm \sim 0.37$ keV. Only two galaxies from \citet{Tullman2006} have some observed regions that are consistent with the range of temperatures seen in our warm component. However, those papers specifically extracted CGM regions separately from the disks of the observed galaxies, so they might have missed any hot component that is localized more closely to the disk.

For both the northern and southern CGM, a trend towards higher temperatures and larger EM for the warm component near the Galactic core can be seen (see Figure 9, also Figures 5 and 6). The hot component doesn't have any obvious core enhancement in temperature, but does exhibit an enhancement in EM. The NPS stands out in the northern CGM hot component maps as a consistently cooler feature with high EM, while the Eridanus Enhancement stands out in the southern CGM warm EM map, but not in the hot EM map nor the temperature maps. The NPS warm component temperature seems typical of neighboring fields, while being much more noticeable in EM.

The wide distribution of EM values (see Figures 4, 6, and 9) between fields points towards the warm component being clumpy in distribution. The hot component also exhibits this wide range of EM, and thus is similarly clumpy. This supports the interpretation of \citet{Kaaret2020} that the CGM has a clumpy distribution, and is consistent with the localized distribution of the hot component in \citet{Hodges2018}.

We can also estimate an angular correlation length for the EM using the same method as \citet{Kaaret2020}. \citet{Kaaret2020} used a model for the CGM distribution and calculated the autocorrelation using the EM difference from that model. We instead compare to the median EM, as we have not generated a model for the two temperature case. This results in correlation lengths for the full northern hemisphere of $\rm 29^{\circ} \pm 4^{\circ}$ for the warm component and $\rm 27^{\circ} \pm 4^{\circ}$ for the hot component. The correlation lengths for the full southern hemisphere are $\rm 18^{\circ} \pm 2^{\circ}$ and $\rm 14^{\circ} \pm 2^{\circ}$ for the warm and hot components, respectively. These correlation lengths are based on the full data set (including the NPS), and reflect the overall trend for EM seen in Figure 9. These errors are for the 90\% confidence interval.

We've previously noted differing properties of fields in the two quadrants closer to the Galactic center versus the two quadrants away from the center. This makes it interesting to examine the correlation length for those subsets. Each data set uses its own median value. We find that the inner quadrants in northern CGM have correlation lengths of $\rm 19^{\circ} \pm 4^{\circ}$ for the hot component and $\rm 11^{\circ} \pm 2^{\circ}$ for the warm component. The southern temperature components have consistent lengths, with $\rm 16^{\circ} \pm 3^{\circ}$ and $\rm 14^{\circ} \pm 5^{\circ}$ for the hot and warm components, respectively. Given that the NPS is such a large difference for the north, different results for each hemisphere is not surprising. The outer two quadrants are consistent for both the northern and southern CGM, an expected result. The warm component for the outer quadrants has a correlation length of $\rm 9^{\circ} \pm 3^{\circ}$ for the north and $\rm 8^{\circ} \pm 3^{\circ}$ for the south. This is consistent with the correlation length of $\rm \sim 6^{\circ}$ for the south found in \citet{Kaaret2020}, in which large scale features were removed by the model. We find that the hot EM values in the outer quadrants with non-zero separation are uncorrelated. This means that the correlation length is less than the $\rm 5^{\circ}$ full response field-of-view of HaloSat. These errors are also for the 90\% confidence interval.

The temperature of the hot component is high enough that it is likely not gravitationally bound to the Galaxy \citep{Spitzer1956}. This would imply that the CGM hot component is very dynamic, and as such the CGM may not be in a steady state. The analytic models of \citet{Pezzulli2017} suggest that the the hot component is driven by either thermal feedback or mechanical feedback (ejected material) from the disk. The hot component would then be produced in star-forming regions before escaping and spreading out into the CGM. Perhaps the observations of external galaxies that point towards the hot component being localized near the disk are only catching the youngest and densest regions of the hot component, near the areas producing the component. While we have ruled against the warm component having an enhancement in neon abundance instead of the CGM having a hot component (as suggested by \citet{Gupta2020}), the work of \citet{Das2019a} points towards the hot component having an enhanced metallicity caused by core-collapse supernovae. {A link to localized star formation regions as a source for the CGM gas may explain the increase in our observed EM towards the Galactic center.} The results presented in this paper are consistent with the interpretation of the hot component being caused by local processes in the disk, such as supernovae, which is a favorable explanation for the observed clumpy distribution of the CGM.

The two temperature model represents the next logical step on expanding from a one temperature model, and is motivated by previous literature \citep{Gupta2020,Das2019c}. However, the particular temperature values found using the two temperature model may not indicate the presence of gas at those specific temperatures (see \citet{KuntzSnowden2010}). The spectral fits may be a simplification of a more complex physical situation with gas distributed over a range of temperatures. In that case, the specific temperature recovered with a two component plasma model may depend on the energy band used for the spectral fitting and how the plasma is sampled along the selected lines of sight. As noted in our previous comparison with the results of \citet{Kaaret2020}, use of a two temperature versus single temperature model shifts the temperatures. Using ROSAT, \citet{KuntzSnowden2000} and \citet{Kuntz2000}, find temperatures of $\rm \sim 0.099$ keV and $\rm \sim 0.25$ keV. These results are not actually inconsistent due to the difference in the energy range of ROSAT versus HaloSat. HaloSat is unable to rigorously detect the softest component seen by ROSAT, while similarly ROSAT cannot rigorously detect the hot component we've seen. As such, the difference in temperatures of the overlapping component would be due to that component covering excess emission from the respective missing third component in each fit. This could explain why the temperature of the warm component is higher in the ROSAT study than in any of the HaloSat studies. It appears there may actually be at least 3 temperature components in the CGM if one looks at a wider energy range, although it is likely that these components are just peaks of a broader distribution of gas temperatures. Some examples of alternative models that could be investigated in the future can be found in \citet{Wang2021} and \citet{Gayley2014}.

If the two temperatures we have detected are a simplified representation of a range of temperatures, then the properties of the two components should be correlated. Figure 10 compares the EM of the warm component to the EM of the hot component. The best fitted linear relation between the components has a slope of $\rm 0.092 \pm 0.017$ and an intercept of $\rm 0.00026 \pm 0.00022$. The correlation coefficient for the data is $\rm 0.73^{+0.08}_{-0.11}$ (90\% confidence interval). Since the intercept is so close to zero, the slope is effectively the ratio of the component EMs. The linear trend persists if harsher or looser restrictions are applied to the included fields, with the slope maxing out around 0.13 for the full data set, and dropping towards $\rm \sim 0.06$ as more high EM fields are removed. The slopes for the north and south data sets are consistent, even though the distribution of the north and south data visually appears a bit different. Our EM ratio of 0.092 is 2-3 time higher than the EM ratio for NGC 891 from \citet{Hodges2018} of $\rm \sim 0.03$. 

The observed correlation shows that the hot component is not entirely separate from the warm component, and provides a further argument against M stars being a significant contributor to the hot component. These components are likely representative of a wider spread of actual temperatures in the CGM. The components could simply be co-spatial, but that situation seems physically unlikely as it would not be stable. Instead, perhaps the hot component cools into the warm component as it moves away from the Galactic disk. Another possibility is that the hot component and the warm component are not directly related. Both components could stem from star formation in regions of the Galactic disk and thus scale in tandem with SFR, with separate regions of both warm and hot emission along the line of sight. Perhaps the truth is a range of temperatures produced by star formation feedback, cooling into each other over time, with the CGM exhibiting a wider range of CGM temperatures than we are currently able to detect. Future instrumentation may help resolve this dilemma. A grating instrument with spectral resolution R $\rm >2500$ \citep{Smith2019}, or a wide-field microcalorimeter instrument with 2 eV spectral resolution, could definitely resolve the issue. Ultimately, reality is likely (and unsurprisingly) more complex than our models.


Our results have revealed the structure of the X-ray emission of the CGM in unprecedented detail, refining the X-ray spectral model of the CGM. Furthermore, our results serve to bring emission studies of our Galaxy more in line with observations of external galaxies, as well as bridge X-ray band gaps in existing and previous X-ray observatories. A catalog of the spectral model values for the northern and southern CGM will be made available at the VizieR Online Data Catalog.

\section*{Acknowledgements}
This research was supported by NASA grants NNX15AU57G, 80NSSC20K0398, and 80NSSC22K0624. D.K.'s HSWCX modeling work was supported by CNES and performed with the High Performance Computer and Visualisation platform (HPCaVe) hosted by UPMC - Sorbonne Université. We'd like to thank Joel Bregman and the anonymous AAS statistician for their helpful comments.
\software{linmix https://github.com/jmeyers314/linmix, matplotlib \citep{Hunter2007}, NumPy \citep{Harris2020}, SciPy \citep{Virtanen2020}, XSPEC (v12.11.1; \citet{Arnaud1996})}

\appendix
\section{Gain correction using spectra of Cas A}
To check the on-orbit X-ray energy scale calibration, we examined spectra obtained while observing the supernova remnant (SNR) Cassiopeia A (Cas A). Cas A is chosen as a calibration target because of the strong emission lines present in its X-ray spectrum \citep{Holt1994,Jahoda2006}. The HaloSat field centered on Cas A includes another SNR, CTB 109, and several point sources, but the emission is dominated by Cas A.

We analyzed all of the observations of Cas A performed during the full HaloSat mission. Data were filtered using selections on the VLE ($\rm > 7$ keV) count rate of 0.75 c/s and on the hard-band (3-7 keV) count rate of 0.12 c/s. We extracted one summed spectrum for each of the three DPUs (Figure 11) with channels grouped to have a minimum of 25 counts per bin. The instrumental background was modelled as a power law with a diagonal response matrix. We performed fits with the power law photon index calculated as described in \citet{Kaaret2020} and with the photon index for each DPU left as a free parameter.  Emission from the cosmic X-ray background was modeled as the sum of an absorbed power law with fixed absorption, photon index, and normalization. The emission from Cas A was modeled as the sum of an absorbed power law with absorption fixed to $\rm 1.47 x 10^{21}$ $\rm cm^{-2}$ and a set of nine, narrow Gaussian emission lines. The set of lines used is the same as in \citet{Kaaret2019}.

\begin{figure}
\centering
\includegraphics[width=0.75\textwidth]{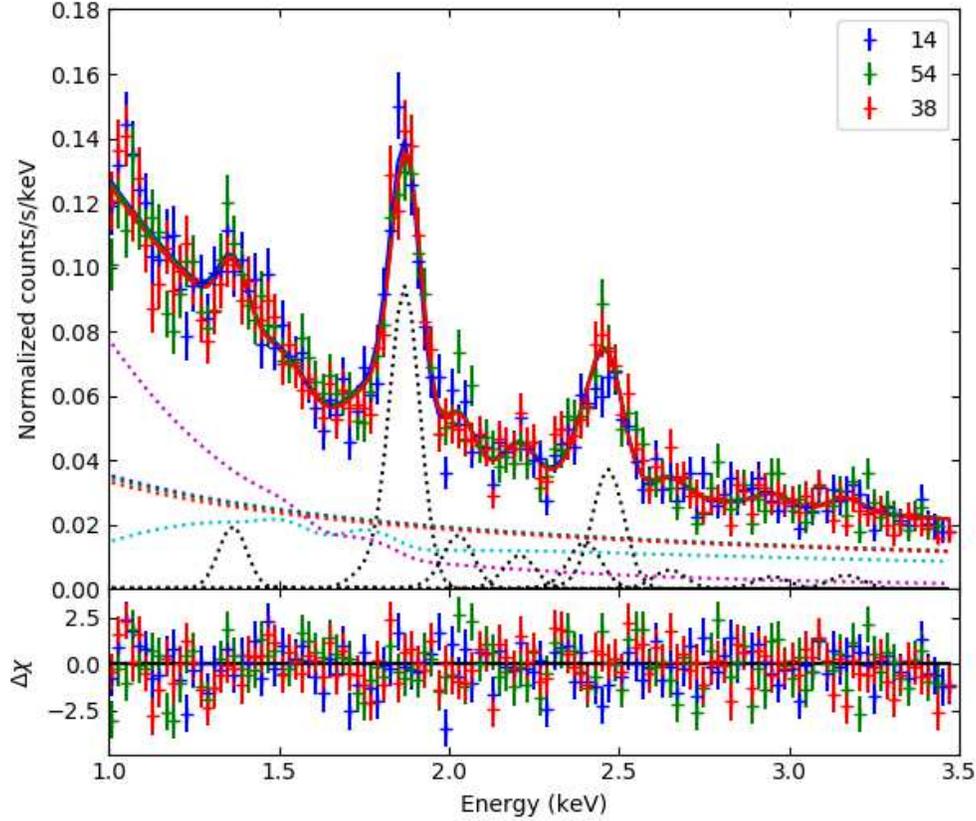}
\caption{X-ray spectra of Cas A. Data are shown for all three DPUs: 14=red, 54=green, 38=red. The fitted instrumental background for each DPU is shown as a dotted curve with the same color as the data. The dotted cyan curve is the cosmic X-ray background. The dotted magenta curve is the power law from Cas A. The dotted black curves are the Gaussian emission lines.} 
\end{figure}

A response model was included in the fitting. Allowing the offset to vary while fixing the slope at unity improved the fit from $\rm \chi^2$/DoF = 824.2/355 to $\rm \chi^2$/DoF = 437.5/352.  Varying the slope while keeping the offset fixed to zero also improved the fit (to $\rm \chi^2$/DoF = 452.9/352), but the model with the varying offset is preferred.  The best fitted model with varying offset is shown in Figure 11.  The best fitted offsets with slope fixed at unity and 90\% confidence uncertainties are given in Table 2. Allowing the instrumental background photon index to vary did not significantly improve the fit, giving $\rm \chi^2$/DoF = 433.1/349 corresponding to an F-test probability of 0.32. There was no significant change in the best fitted offsets (Table 3). Averaging the values and rounding to 1 eV, leads to an offset of 0.0232 keV for DPU 14, 0.0240 keV for DPU 54, and 0.0239 keV for DPU 38. These offsets were tested using other HaloSat data sets.

\begin{deluxetable}{cccc}
\tablenum{2}
\tablecaption{Response Parameters\label{tab:MP}}
\tablewidth{0pt}
\tablehead{
\colhead{DPU} & \colhead{offset} & \colhead{lower error} & \colhead{upper error}\\
}
\startdata
14 & 0.02320 & -0.00191 & +0.00074\\
54 & 0.02402 & -0.00065 & +0.00057\\
38 & 0.02393 & -0.00063 & +0.00054\\
\enddata
\tablecomments{Response parameters using calculated background slope. Units are keV.}
\end{deluxetable}

\begin{deluxetable}{cccc}
\tablenum{3}
\tablecaption{Response Parameters\label{tab:MP}}
\tablewidth{0pt}
\tablehead{
\colhead{DPU} & \colhead{offset} & \colhead{lower error} & \colhead{upper error}\\
}
\startdata
14 & 0.02316 & -0.00181 & +0.00075\\
54 & 0.02398 & -0.00067 & +0.00058\\
38 & 0.02389 & -0.00064 & +0.00055\\
\enddata
\tablecomments{Response parameters using fitted background slope. Units are keV.}
\end{deluxetable}

The new gain offset was tested using the southern CGM data from \citet{Kaaret2020}. Application of the gain offsets in Table 1 lead to: 1) No significant change in the Cash statistic. Median changed from 1.067 to 1.062. 2) A small change in the best fitted temperatures. The median change in kT is -0.0196 keV and the weighted average change is -0.0188 keV. 3) A moderate change in the best fitted emission measure (EM). The median of the fraction change in EM is 15\% and 96\% of fields have a fraction change in EM of less than 30\%. The median of the change in EM divided by the statistical error on the EM is 0.67.

Spectra were produced by summing a large number of CGM fields with consistent emission measure. Processing and background estimation were performed as described above.  The spectra were fitted with a model consisting of an APEC for the local hot bubble, an absorbed power law for the cosmic X-ray background, and two absorbed APEC models for the CGM emission.

Figure 12 shows the model fits with the original HaloSat gain values. Figure 13 shows the model fits with the gain offsets applied as listed in Table 2.  All spectra are rebinned for presentation using setplot rebin 30 15, for 30 sigma per bin with a cap of 15 channels combined. 

\begin{figure}
\centering
\includegraphics[width=0.75\textwidth]{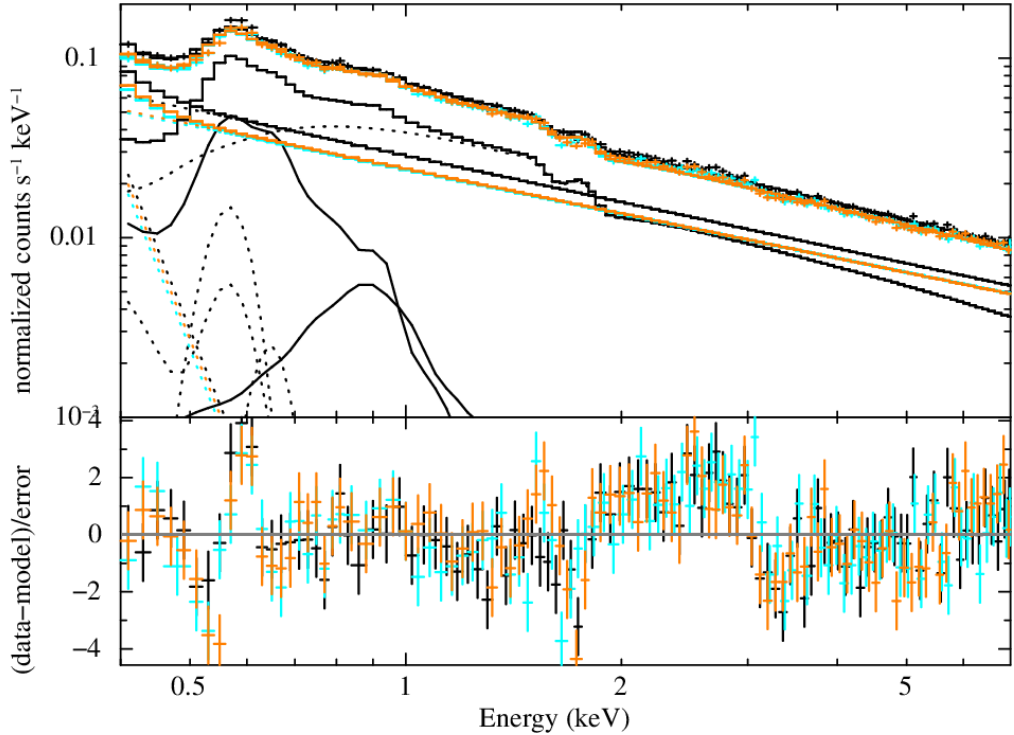}
\caption{Stacked CGM spectra before gain correction.} 
\end{figure}

\begin{figure}
\centering
\includegraphics[width=0.75\textwidth]{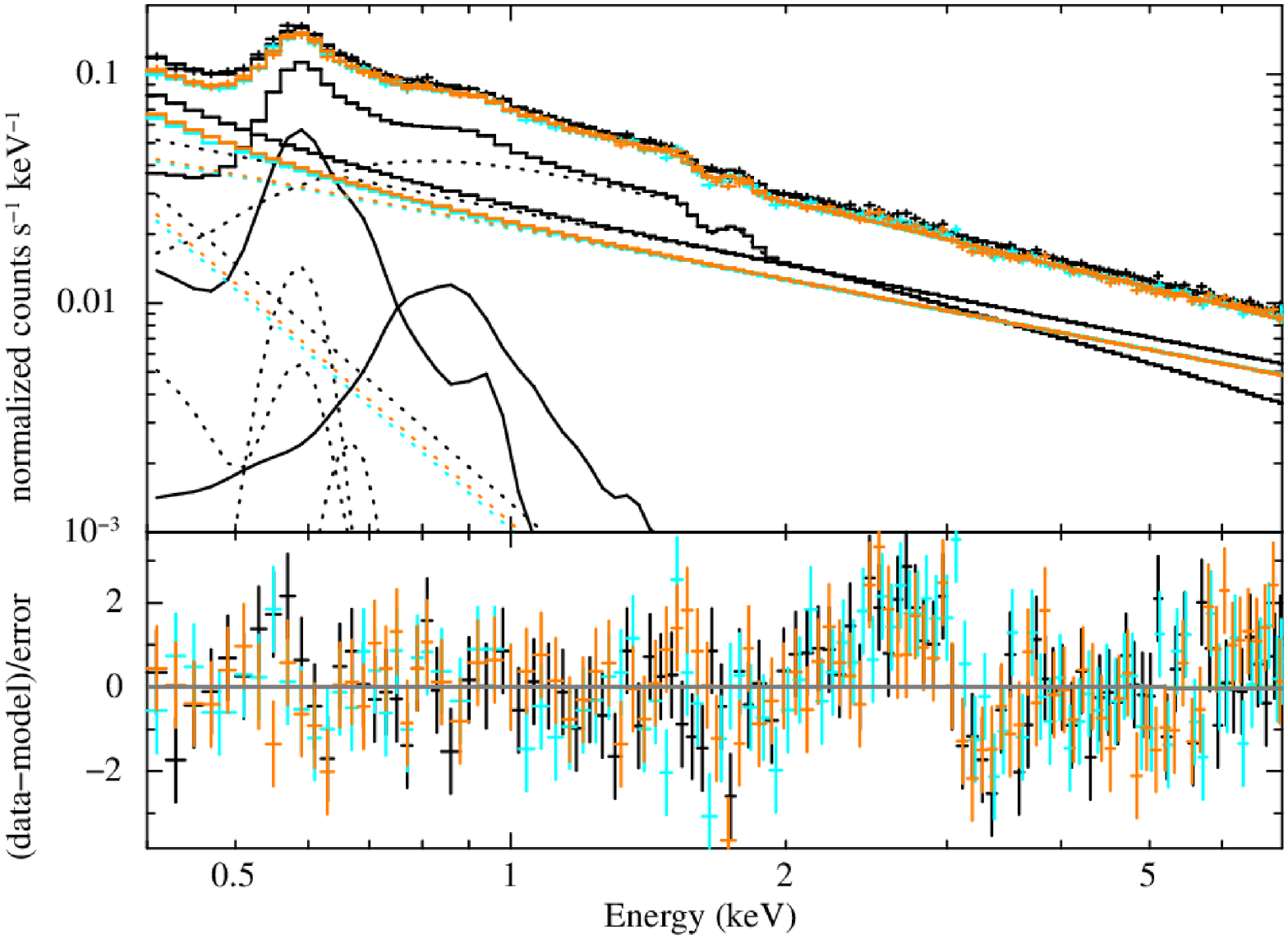}
\caption{Stacked CGM spectra after gain correction. Note that this also includes the edge component discussed in Appendix B.} 
\end{figure}

The gain corrections significantly improve the residuals of the fit. The fit statistic improved from $\rm \chi^2$/DoF = 1266/973 to $\rm \chi^2$/DoF = 1069/973.  The gain corrections produced a noticeable improvement from ~0.45 keV to ~0.8 keV. This had a significant effect on the model – the CGM temperatures before the gain are inconsistent with those found after.

We recommend that anyone using the HaloSat response files released on 2020-03-20 add to their analysis an offset of 0.0232 keV for DPU 14, 0.0240 keV for DPU 54, and 0.0239 keV for DPU 38. The gain offset appears to have a modest effect on spectra with low statistics fitted with models with a single free component. However, the gain offset appears to be more consequential for spectra with high statistics and/or fits with multiple components. 

\section{Investigation of edge features in the response using spectra of the Crab}

To check incorrectly modeled edge features in the HaloSat response matrix, we examined spectra obtained while observing the Crab pulsar wind nebula and pulsar. The Crab is chosen for this purpose because it has a featureless X-ray spectrum. The emission within the HaloSat field centered on the Crab is dominated by the Crab.

We analyzed all of the observations of the Crab performed during the full HaloSat mission. Data were filtered using selections on the VLE ($\rm > 7$ keV) count rate of 0.75 c/s and on the hard band (3-7 keV) count rate of 0.5 c/s. A Sun angle greater than $\rm 100^{\circ}$ was required. The count rate threshold in the hard band is higher than typically used in HaloSat analysis because the Crab contributes significant flux in the hard band. We note that the feature around 3 keV induced by the standard hard rate selections is not present in the Crab spectra. We extracted one summed spectrum for each of the three DPUs (Figure 11) with channels grouped to have a minimum of 25 counts per bin. The gain corrections described in the previous section were applied. The instrumental background was modelled as a power law with a diagonal response matrix. The normalization and photon index for each DPU left as free parameters. The instrumental background photon index calculation described in \citet{Kaaret2020} are not applicable due to the non-standard data filtering required for the Crab. Emission from the cosmic X-ray background (CXB) was modeled as an absorbed power law with fixed absorption, photon index, and normalization. Emission from the local hot bubble (LHB) was modeled as an unabsorbed APEC. The parameters for the CXB and LHB were set following \citet{Kaaret2020}. The emission from the Crab was modeled as an absorbed power law. The absorption column density and power law photon index and normalization were allowed to vary.

\begin{figure*}[htb!]
\centering
\begin{tabular}{cc}
  \includegraphics[width=0.49\textwidth]{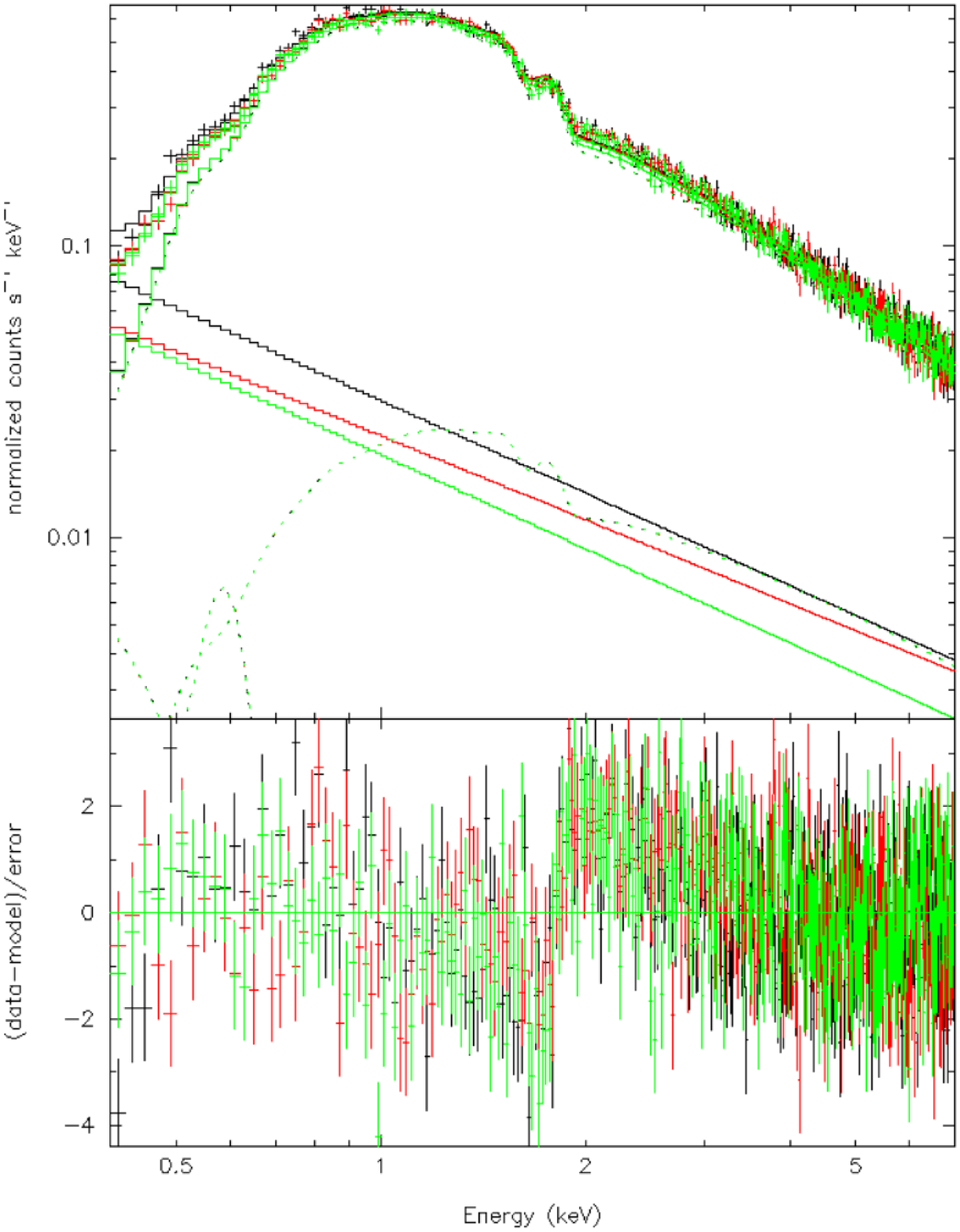} & \includegraphics[width=0.49\textwidth]{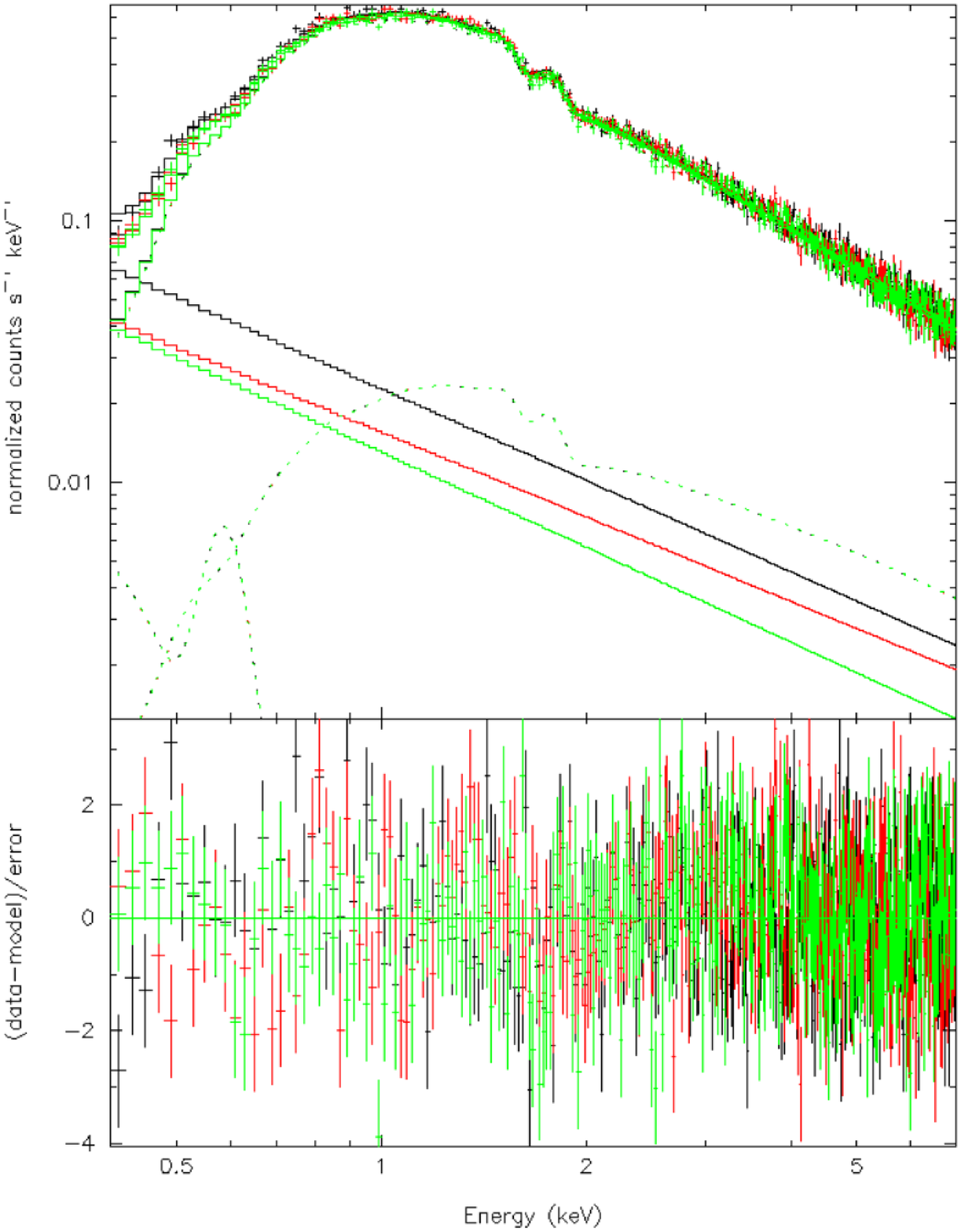} \\
\end{tabular}
\caption{Crab spectrum before (left) and after (right) the inclusion of the negative Si edge. There are initially strong residuals around 1.8 keV. After the inclusion of a negative Si edge, the residuals around 1.8 keV are greatly improved.}
\end{figure*}

Figure 14 (left) shows the spectra for all three detectors. The spectral fits show strong residuals near 1.8 keV which we identify as the silicon K edge. The fit has $\rm \chi^2$/DoF = 1424.38/978. Figure 14 (right) also shows the spectral fits including an edge (model ‘edge’ in Xspec) with the edge energy fixed to 1.839 keV for silicon. The maximum optical depth (tau) is linked between the detectors and the best fit value is tau = -0.169±0.014. Addition of the edge significantly improves the fit to $\rm \chi^2$/DoF = 1084.85/977. The corresponding F-test probability is $\rm 9x10^{-60}$. Allowing tau to vary between detectors produces no significant improvement in the fit (F-test probability = 0.38). The error ranges on the tau values for the individual detectors all include the linked fit value of tau = -0.169.

We recommend inclusion of a negative absorption edge using the Xspec edge model with a threshold energy of 1.839 keV and tau of 0.169 in all model fits when using the HaloSat response files released on 2020-03-20. We note that the edge is ‘negative’. This means that there is less absorption than included in the current response model. An Si thickness of 0.24 microns produces a step in transmission across the Si K edge equivalent to the measured tau. Such a change in Si thickness should also affect the transmission at low energies, but this is not observed. The equivalent Si thickness is comparable to the thickness of front layer (0.11 microns) and incomplete charge collection layer (0.20 microns) of the HaloSat silicon drift detectors \citep{Zajczyk2020}. The edge may result from incomplete modeling of the interactions of the Auger and photo-electrons in those regions \citep{Scholze2009}.


\begin{thebibliography}{}
\bibitem[Arnaud(1996)]{Arnaud1996} Arnaud, K. A. 1996, in Astronomical Data Analysis Software and Systems V, ed. J. H. Jacoby \& J. Barnes (San Francisco : ASP), 17
\bibitem[Bigiel et al.(2008)]{Bigiel2008}Bigiel, F., Leroy, A., Walter, F., Brinks, E., de Blok, W.~J.~G., Madore, B., Thornley, M.~D., 2008, AJ, 136, 2846
\bibitem[Bland-Hawthorn \& Maloney(1999)]{Bland1999} Bland-Hawthorn, J., \& Maloney, P. R., 1999, ApJL, 510, L33
\bibitem[Blitz \& Robishaw(2000)]{Blitz2000} Blitz, L., \& Robishaw, T., 2000, ApJ, 541, 675
\bibitem[Bregman \& Lloyd-Davies(2007)]{Bregman2007b} Bregman, J.~N., Lloyd-Davies, E.~J., 2007, ApJ, 669, 990
\bibitem[Bregman et al.(2018)]{Bregman2018} Bregman, J.~N., Anderson, M.~E., Miller, M.~J., Hodges-Kluck, E., Dai, X., Li, J.-T., Li, Y., et al., 2018, ApJ, 862, 3
\bibitem[Br{\"u}ns et al.(2000)]{Bruns2000} Br{\"u}ns, C., Kerp, J., Kalberla, P.~M.~W., \& Mebold, U., 2000, A\&A, 357, 120
\bibitem[Burrows et al.(1993)]{Burrows1993}Burrows, D.~N., Singh, K.~P., Nousek, J.~A., Garmire, G.~P., Good, J., 1993, ApJ, 406, 97
\bibitem[Cappelluti et al.(2017)]{Cappelluti2017} Cappelluti, N., Li, Y., Ricarte, A., et al.\ 2017, ApJ, 837, 19
\bibitem[Das et al.(2019a)]{Das2019a} Das, S., Mathur, S., Nicastro, F., \& Krongold, Y., 2019a, ApJL, 882, L23
\bibitem[Das et al.(2019b)]{Das2019b} Das, S., Mathur, S., Gupta, A., Nicastro, F., Krongold, Y., Null, C., 2019, ApJ, 885, 108
\bibitem[Das et al.(2019c)]{Das2019c} Das, S., Mathur, S., Gupta, A., Nicastro, F., \& Krongold, Y., 2019c, ApJ, 887, 2
\bibitem[Faerman et al.(2017)]{Faerman2017} Faerman, Y., Sternberg, A., McKee ,C.~F., 2017, ApJ, 835, 52
\bibitem[Gayley(2014)]{Gayley2014}Gayley, K.~G., 2014, ApJ, 788, 90
\bibitem[Gloeckler et al.(1998)]{Gloeckler1998}Gloeckler, G., Cain, J., Ipavich, F.~M., Tums, E.~O., Bedini, P., Fisk, L.~A., Zurbuchen, T.~H., et al., 1998, SSRv, 86, 497
\bibitem[Goodman \& Weare(2010)]{Goodman2010}Goodman, J., Weare, J., 2010, CAMCS, 5, 65
\bibitem[Gordon \& Arnaud(2021)]{Gordon2021}Gordon, C., Arnaud, K., 2021, ascl.soft. ascl:2101.014
\bibitem[Grcevich \& Putman(2009)]{Grcevich2009} Grcevich, J., \& Putman, M. E., 2009, ApJ, 696, 385
\bibitem[Gupta et al.(2012)]{Gupta2012} Gupta, A., Mathur, S., Krongold, Y., Nicastro, F., Galeazzi, M., 2012, ApJL, 756, L8
\bibitem[Gupta et al.(2014)]{Gupta2014} Gupta, A., Mathur, S., Galeazzi, M., Krongold, Y., 2014, Ap\&SS, 352, 775
\bibitem[Gupta et al.(2021)]{Gupta2020} Gupta, A., Kingsbury, J., Mathur, S., Das, S., Galeazzi, M., Krongold, Y., Nicastro, F., 2021, ApJ, 909, 164
\bibitem[Harris et al.(2020)]{Harris2020} Harris, C.R., Millman, K.J., van der Walt, S.J. et al., 2020, Nature, 585, 357
\bibitem[Heckman \& Borthakur(2016)]{Heckman2016} Heckman, T. M., \& Borthakur, S., 2016, ApJ, 822, 9
\bibitem[Henley \& Shelton(2013)]{Henley2013} Henley, D.~B., Shelton, R.~L., 2013, ApJ, 773, 92
\bibitem[Hodges-Kluck, Bregman, \& Li(2018)]{Hodges2018} Hodges-Kluck, E.~J., Bregman, J.~N., Li, J.-. tao ., 2018, ApJ, 866, 126
\bibitem[Holt et al.(1994)]{Holt1994}Holt, S.~S., Gotthelf, E.~V., Tsunemi, H., Negoro, H., 1994, PASJ, 46, L151
\bibitem[Hunter(2007)]{Hunter2007}Hunter, J. D. 2007, CISE, 9(3), 90
\bibitem[Immler et al.(2003)]{Immler2003} Immler, S., Wang, Q.~D., Leonard, D.~C., Schlegel, E.~M., 2003, ApJ, 595, 727
\bibitem[Jahoda et al.(2006)]{Jahoda2006} Jahoda, K., Markwardt, C.~B., Radeva, Y., Rots, A.~H., Stark, M.~J., Swank, J.~H., Strohmayer, T.~E., et al., 2006, ApJS, 163, 401
\bibitem[Joye \& Mandel(2003)]{Joye2003} Joye, W.~A. \& Mandel, E., 2003, ASPC, 295, 489
\bibitem[Kaaret et al.(2019)]{Kaaret2019} Kaaret, P., Zajczyk, A., LaRocca, D., et al.\ 2019, ApJ, 884, 162
\bibitem[Kaaret et al.(2020)]{Kaaret2020} Kaaret, P., Koutroumpa, D., Kuntz, K. D., et al.\ 2020, NatAs, 4, 1072
\bibitem[Kelly(2007)]{Kelly2007}Kelly, B.~C., 2007, ApJ, 665, 1489
\bibitem[Klypin et al.(2002)]{Klypin2002} Klypin, A., Zhao, H., \& Somerville, R.~S., 2002, ApJ, 573, 597
\bibitem[Konz(2002)]{Konz2002}Konz, C., Br{\"u}ns, C., Birk, G.~T., 2002, A\&A, 391, 713
\bibitem[Koutroumpa et al.(2007)]{Koutroumpa2007}Koutroumpa, D., Acero, F., Lallement, R., Ballet, J., Kharchenko, V., 2007, A\&A, 475, 901
\bibitem[Kuntz(2000)]{Kuntz2000} Kuntz, K.~D., 2000, PhDT
\bibitem[Kuntz \& Snowden(2000)]{KuntzSnowden2000} Kuntz, K.~D., Snowden, S.~L., 2000, ApJ, 543, 195
\bibitem[Kuntz et al.(2003)]{Kuntz2003}Kuntz K.~D., Snowden S.~L., Pence W.~D., Mukai K., 2003, ApJ, 588, 264
\bibitem[Kuntz \& Snowden(2010)]{KuntzSnowden2010} Kuntz, K.~D., Snowden, S.~L., 2010, ApJS, 188, 46
\bibitem[Kuntz(2019)]{Kuntz2019} Kuntz, K.~D., 2019, A\&ARv, 27, 1
\bibitem[LaRocca et al.(2020a)]{LaRocca2020} LaRocca, D. M., Kaaret, P., Kirchner, D. L., et al.\ 2020, JATIS, 6, 014003
\bibitem[LaRocca et al.(2020b)]{LaRocca20202} LaRocca, D.~M., Kaaret, P., Kuntz, K.~D., Hodges-Kluck, E., Zajczyk, A., Bluem, J., Ringuette, R., et al., 2020, ApJ, 904, 54
\bibitem[Li et al.(2008)]{Li2008} Li, J.-T., Li, Z., Wang, Q.~D., Irwin, J.~A., Rossa J., 2008, MNRAS, 390, 59
\bibitem[Liu et al.(2017)]{Liu2017} Liu, W., Chiao, M., Collier, M. R., et al.\ 2017, ApJ, 834, 33
\bibitem[Masui et al.(2009)]{Masui2009} Masui K., Mitsuda K., Yamasaki N.~Y., Takei Y., Kimura S., Yoshino T., \& McCammon D., 2009, PASJ, 61, S115
\bibitem[McCammon et al.(2002)]{McCammon2002} McCammon, D., Almy, R., Apodaca, E., Bergmann Tiest, W., Cui, W., Deiker, S., Galeazzi, M., et al., 2002, ApJ, 576, 188
\bibitem[McMillan(2011)]{McMillan2011} McMillan, P.~J., 2011, MNRAS, 414, 2446
\bibitem[Miller \& Bregman(2015)]{Miller2015} Miller, M.~J., Bregman, J.~N., 2015, ApJ, 800, 14
\bibitem[Mitsuishi et al.(2012)]{Mitsuishi2012} Mitsuishi, I., Gupta, A., Yamasaki, N.~Y., Takei, Y., Ohashi, T., Sato, K., Galeazzi, M., et al., 2012, PASJ, 64, 18
\bibitem[Muller et al.(1963)]{Muller1963} Muller, C.~A., Oort, J.~H., \& Raimond, E., 1963, CRAS, 257, 1661
\bibitem[Nakashima et al.(2018)]{Nakashima2018} Nakashima, S., Inoue, Y., Yamasaki, N., Sofue, Y., Kataoka, J., Sakai, K., 2018, ApJ, 862, 34
\bibitem[Nicastro et al.(2016)]{Nicastro2016}Nicastro, F., Senatore, F., Krongold, Y., Mathur, S., Elvis, M., 2016, ApJL, 828, L12
\bibitem[Oser et al.(2010)]{Oser2010} Oser, L., Ostriker, J.~P., Naab, T., Johansson, P.~H., Burkert, A., 2010, ApJ, 725, 2312
\bibitem[Owen \& Warwick(2009)]{Owen2009} Owen, R.~A., Warwick, R.~S., 2009, MNRAS, 394, 1741
\bibitem[Pezzulli et al.(2017)]{Pezzulli2017}Pezzulli, G., Fraternali, F., Binney, J., 2017, MNRAS, 467, 311
\bibitem[Planck Collaboration et al.(2014)]{Planck2014} Planck Collaboration, et al., 2014, A\&A, 571, A11
\bibitem[Putman et al.(2012)]{Putman2012} Putman, M.~E., Peek, J.~E.~G., \& Joung, M.~R., 2012, ARA\&A, 50, 491
\bibitem[Ringuette et al.(2021)]{Ringuette2021}Ringuette, R., Koutroumpa, D., Kuntz, K.~D., Kaaret, P., Jahoda, K., LaRocca, D., Kounkel, M., et al., 2021, ApJ, 918, 41
\bibitem[Salem et al.(2015)]{Salem2015} Salem, M., Besla, G., Bryan, G., Putman, M., van der Marel, R.~P., Tonnesen, S., 2015, ApJ, 815, 77
\bibitem[Scholze \& Procop(2009)]{Scholze2009} Scholze, F., Procop, M., 2009, XRS, 38, 312
\bibitem[Sembach et al.(2003)]{Sembach2003} Sembach, K. R., Wakker, B. P., Savage, B. D., et al., 2003, ApJS, 146, 165
\bibitem[Smith et al.(2019)]{Smith2019} Smith, R.~K., Abraham, M., Baird, G., Bautz, M., Bookbinder, J., Bregman, J., Brenneman, L., et al., 2019, SPIE, 11118, 111180W
\bibitem[Smith et al.(2001)]{Smith2001} Smith, R.~K., Brickhouse, N.~S., Liedahl, D.~A., \& Raymond, J.~C., 2001, ApJL, 556, L91
\bibitem[Snowden et al.(1995)]{Snowden1995} Snowden, S.~L., Burrows, D.~N., Sanders, W.~T., Aschenbach, B., \& Pfeffermann, E., 1995, ApJ, 439, 399
\bibitem[Snowden et al.(1998)]{Snowden1998} Snowden, S.~L., Egger, R., Finkbeiner, D.~P., Freyberg, M.~J., Plucinsky, P.~P., 1998, ApJ, 493, 715
\bibitem[Snowden et al.(2000)]{Snowden2000} Snowden, S.~L., Freyberg, M.~J., Kuntz, K.~D., Sanders, W.~T., 2000, ApJS, 128, 171
\bibitem[Sommer-Larsen(2006)]{Sommer2006} Sommer-Larsen, J., 2006, ApJL, 644, L1
\bibitem[Spitzer(1956)]{Spitzer1956} Spitzer, L., 1956, ApJ, 124, 20
\bibitem[Strickland et al.(2004)]{Strickland2004} Strickland, D.~K., Heckman, T.~M., Colbert, E.~J.~M., Hoopes, C.~G., Weaver, K.~A., 2004, ApJS, 151, 193
\bibitem[Su et al.(2010)]{Su2010} Su, M., Slatyer, T. R., \& Finkbeiner, D. P., 2010, ApJ, 724, 2
\bibitem[T{\"u}llmann et al.(2006)]{Tullman2006} T{\"u}llmann, R., Pietsch, W., Rossa, J., Breitschwerdt, D., Dettmar R.-J., 2006, A\&A, 448, 43
\bibitem[Tumlinson et al.(2017)]{Tumlinson2017} Tumlinson, J., Peeples, M.~S., Werk, J.~K., 2017, ARA\&A, 55, 389
\bibitem[Veilleux et al.(2005)]{Veilleux2005} Veilleux, S., Cecil, G., \& Bland-Hawthorn, J., 2005, ARA\&A, 43, 769
\bibitem[Virtanen et al.(2020)]{Virtanen2020} Virtanen, P., Gommers, R., Oliphant, T.~E., Haberland, M., Reddy, T., Cournapeau, D., Burovski, E., et al., 2020, NatMe, 17, 261
\bibitem[Wang et al.(2003)]{Wang2003}Wang, Q.~D., Chaves, T., Irwin, J.~A., 2003, ApJ, 598, 969
\bibitem[Wang et al.(2021)]{Wang2021} Wang, Q.~D., Zeng, Y., Bogd{\'a}n, {\'A}., Ji, L., 2021, MNRAS, 508, 6155
\bibitem[Wilms et al.(2000)]{Wilms2000} Wilms, J., Allen, A., \& McCray, R., 2000, ApJ, 542, 914
\bibitem[Wulf et al.(2019)]{Wulf2019} Wulf, D., Eckart, M.~E., Galeazzi, M., Jaeckel, F., Kelley, R.~L., Kilbourne, C.~A., Morgan, K.~M., et al., 2019, ApJ, 884, 120
\bibitem[Yoshino et al.(2009)]{Yoshino2009} Yoshino, T., Mitsuda, K., Yamasaki, N.~Y., Takei, Y., Hagihara, T., Masui, K., Bauer, M., et al., 2009, PASJ, 61, 805
\bibitem[Zajczyk et al.(2020)]{Zajczyk2020} Zajczyk, A., Kaaret, P., LaRocca, D., et al., 2020, , JATIS, 6, 044005
\bibitem[Zhu et al.(2017)]{Zhu2017} Zhu, H., Tian, W., Li, A., \& Zhang, M.,\ 2017, MNRAS, 471, 3494
\end{thebibliography}
{}


\end{document}